\documentclass[iop]{emulateapj}
\slugcomment{Accepted for publication: The Astrophysical Journal, 2016, February 3}

\newcommand{\feii}{Fe\,{\sc ii}}
\newcommand{\oi}{O\,{\sc i}}
\newcommand{\caii}{Ca\,{\sc ii}}
\newcommand{\hb}{H$\beta$}
\newcommand{\paa}{Pa$\alpha$}
\newcommand{\pab}{Pa$\beta$}

\newcommand{\lya}{Ly$\alpha$}
\newcommand{\mum}{$\mu$m}
\newcommand{\Rone}{R$_{1\mu m}$}
\newcommand{\Rfour}{R$_{4570}$}
\newcommand{\Rnine}{R$_{9200}$}
\shorttitle{The \feii\ emission in AGNs.}
\shortauthors{Marinello et al.}

\begin{document}
\title{The \feii\ emission in active galactic
	nuclei:\\excitation mechanisms and location of the emitting region}
\author{A. O. M. Marinello\altaffilmark{1}} 
\affil{Universidade Federal de Itajub\'a, 
	 Rua Doutor Pereira Cabral 1303, 37500-903, 
	 Itajub\'a, MG, Brazil}
\email{murilo.marinello@gmail.com}
	 
\author{A. Rodr\'{\i}guez-Ardila\altaffilmark{2} 
	 and A. Garcia-Rissmann}
\affil{Laborat\'orio Nacional de Astrof\'isica, Rua Estados Unidos 154,
	Itajub\'a, MG, 37504-364, Brazil}

\author{T. A. A. Sigut\altaffilmark{2}}
\affil{The University of Western Ontario, 
	London, Ontario N6A 3K7, Canada}

\and

\author{A. K. Pradhan}
\affil{McPherson Laboratory, The Ohio State 
	University, 140 W. 18th Ave., Columbus, 
	OH 43210-1173, USA}

\altaffiltext{1}{Present address: Observat\'orio 
	Nacional, Rua Gal Jos\'e Cristino 77, 
	20921-400, Rio de Janeiro, RJ, Brazil}

\altaffiltext{2}{Visiting Astronomer at the Infrared 
	Telescope Facility, which is operated by the 
	University of Hawaii under Cooperative Agreement 
	no. NNX-08AE38A with the National Aeronautics and 
	Space Administration, Science Mission Directorate, 
	Planetary Astronomy Program.}
\begin{abstract}
We present a study of \feii\ emission in the 
  near-infrared region (NIR) for 25 active galactic 
  nuclei (AGNs) to obtain information 
  about the excitation mechanisms that power it 
  and the location where it is formed. 
We employ a NIR \feii\ template 
  derived in the literature and found that it 
  successfully reproduces the observed \feii\ spectrum. 
The \feii\ bump at 9200\AA{} detected
  in all objects studied confirms that \lya\ 
  fluorescence is always present in AGNs. 
The correlation found between the flux of the 9200\AA{} bump, the 1\,\mum\
  lines and the optical \feii\ imply that \lya\ fluorescence plays an 
  important role in the \feii\ production. 
We determined that at least 18$\%$ of 
  the optical \feii\ is due to this process 
  while collisional excitation 
  dominates the production of the observed 
  \feii. 
The line profiles of \feii\,$\lambda$10502, 
  \oi\,$\lambda$11287, 
  \caii\,$\lambda$8664 
  and \pab\ were compared 
  to gather information about the most likely 
  location where they are emitted. 
We found that \feii, \oi\ and 
  \caii\ have similar widths and are, on 
  average, 30$\%$ narrower than \pab.
Assuming that the clouds emitting the lines are 
  virialized, we show that the \feii\ 
  is emitted 
  in a region twice as far from the central source 
  than \pab. 
The distance though strongly varies: from 8.5 
  light-days for NGC\,4051 to 198.2 light-days 
  for Mrk\,509. 
Our results reinforce the importance of 
  the \feii\ in the NIR to 
  constrain critical parameters that drive its physics	 
  and the underlying AGN kinematics as well as more accurate 
  models aimed at reproducing this complex emission.
\end{abstract}
\keywords{galaxies: active --- galaxies: Seyfert --- 
	   infrared: general --- quasars: emission lines --- 
	   techniques: spectroscopic}

\section{Introduction}\label{introduction}
The broad line region (BLR) in active galactic nuclei (AGNs) 
  has been extensively studied from X-rays to the near 
  infrared region (NIR) during the last decades 
  \citep[see][for a review]{gas09} but several aspects 
  about its physical properties remain under debate. 
That is the case of the \feii\ emission, 
  whose numerous multiplets form a pseudo-continuum that 
  extends from the ultraviolet (UV) to the NIR due to the 
  blending of over 344,000 transitions \citep{bru08} 
  although it is not clear that even that number of lines could 
  denote adequate coverage. 
This emission is significant for at least four reasons: 
(\textit{i}) it represents one of the most conspicuous 
  cooling agents of the BLR, emitting about 25\% of the 
  total energy of this region \citep{wil85}; 
(\textit{ii}) it represents a strong contaminant because 
  of the large number of emission lines. 
  Without proper modeling and subtraction, it may lead 
  to a wrong description of the BLR physical conditions; 
(\textit{iii}) the gas responsible for the \feii\ 
  emission can provide important clues on the structure 
  and kinematics of the BLR and the central source. 
  However, despite the extensive study of the 
  \feii\ emission, 
  \citep{kue08,mat08,pop07,slu07,hu08,kov10} 
  the strong blending of the lines prevents an accurate 
  study of its properties and excitation mechanisms; 
(\textit{iv}) The strength of \feii\ relative to 
  the peak of [O\,{\sc iii}], the so called eigenvector 1, 
  which consists of the dominant variable in the principal 
  component analysis presented by \citet{bor92}, 
  is believed to be associated to important parameters 
  of the accretion process \citep{sul00,bor92}.

Due to its complexity and uncertainties in  
  transition probabilities and excitation mechanisms, 
  the most successful approach 
  to model the \feii\ emission in AGNs consists of 
  deriving empirical templates from observations. 
Among the most successful templates for the optical 
  region are the ones of \cite{bor92} 
  and \cite{ver04}, which were developed using 
  the spectrum of I\,Zw\,1, the prototype of NLS1 that is widely 
  known for its strong iron emission in the optical and UV regions 
  \citep{jol91,law97,rud00,tsu06,bru08}.
Other works also successfully employ templates
  to quantify the optical \feii\ emission in larges samples
  of AGNs \citep{tsu06,pop07,kov10,don10,don11}.
For the UV region, \cite{ves01} extended the \feii\ template using 
  Hubble Space Telescope/FOS spectra of I\,Zw\,1,
  presenting the first UV template for this emission.

Two decades after the seminal work by \cite{bor92} on 
  the optical FeII emission template, the first 
  semi-empirical NIR FeII template was derived by 
  \cite{gar12} using a mid-resolution spectrum of 
  I\,Zw\,1 and theoretical models of \cite{sig04} 
  and \cite{sig03}. 
They successfully modeled the  \feii\ emission 
  in that galaxy as well as Ark\,564, another NLS1 
  known for its conspicuous iron emission 
  \citep{jol91,rod02}. 
Similar to the optical, they found that the \feii\ 
  spectrum in the NIR forms a subtle pseudo-continuum 
  that extends from 8300\,\AA{} up to 11800\,\AA{}. 
However, unlike in the optical, the NIR  \feii\ 
  spectrum displays prominent lines that can be 
  fully isolated, allowing the characterization of 
  \feii\ emission line profiles and its comparison 
  to other BLR emission features. 
That property confers an advantage to the NIR over 
  the optical, making the \feii\ emission in that 
  region a powerful tool to study and understand this 
  complex emission. 
It can be used, for instance, to observationally 
  constrain the most likely location of the region 
  emitting this ion. 
Previous works in the NIR carried out on a few targets 
  \citep{rod02} suggest that the \feii\ lines are 
  preferentially formed in the outer part of the BLR. 
Studies on a larger number of sources are 
  necessary to confirm this trend and compare it to 
  results obtained in the optical 
  \citep{kov10,pop07,bor92}. 

Despite the wide and successful use of templates to 
  reproduce, measure and subtract the \feii\ 
  emission in AGNs \citep{bor92,ves01,ver04,gar12}, 
  attempts on determining the mechanisms that drive 
  this emission continue to be an open issue. 
Current models \citep{bal04,ver99,bru08} include processes 
  such as continuum fluorescence, collisional excitation 
  and self-fluorescence among \feii\ transitions. 
  They are successful at reproducing the emission lines
  strengths for the UV and optical lines but results for the 
  NIR are not presented, very likely because the relevant transitions
  in that region are not included.
\cite{pen87} proposed that \lya\ fluorescence could 
  be a key process involved in the production of 
  \feii. 
Indeed, models developed by \cite{sipr98,sig03} and \cite{sig04} showed 
  that this mechanism is of fundamental importance in determining the 
  strength of \feii\ in the NIR. 
The key feature that would reveal the presence of such 
  mechanism is the detection of the \feii\ blend 
  at 9200\,\AA{}, first identified by \cite{rod02} in AGNs. 
The bulk of this emission would be produced by primary 
  cascades from the upper 5p levels to e$^4$D 
  and e$^6$D levels after a capture of a \lya\ photon. 

Additional NIR features resulting from secondary cascading 
  after the \lya\ fluorescence process are 
  the so-called 1\,\mum\ \feii\ lines ($\lambda$9997, 
  $\lambda$10502, $\lambda$10862 and $\lambda$11127), 
  which are the most prominent \feii\ features in the 
  whole 8000-24000~\AA\ region \citep{sig03,pana11}. 
The importance of the 1\,\mum\ lines resides in the fact 
  that they are produced by the decay of the e$^4$D and 
  e$^6$D levels down to the z$^4$D$^0$ and z$^4$F$^0$ levels. 
Transitions downwards from the latter two populate the upper 
  levels responsible for nearly 50\% of all optical 
  \feii\ emission \citep{ver04}. 
If \lya\ fluorescence plays a key role as excitation 
  mechanism of \feii\ in AGNs, a direct correlation 
  should be observed between the strength of 9200\,\AA{} 
  blend and the 1~$\mu$m \feii\ lines in the NIR. 
This issue should be investigated in detail because 
  it can provide useful constrains to the \feii\ 
  problem.

In this paper, we describe for the first time a detailed 
  application of the semi-empirical NIR \feii\ 
  template developed by \cite{gar12} to a sample of 
  25 AGNs. 
The aims are threefold: 
(i) Provide a reliable test for the NIR \feii\ 
  template and verify its capability to reproduce 
  the numerous \feii\ emission lines 
  in a  sample of Type~1 AGNs. 
(ii) Measure the NIR \feii\ flux and compare it 
  to that of the optical region to confirm model 
  predictions of the role of \lya\ fluorescence in 
  the total \feii\ strength. 
(iii) Compare the \feii\ emission line profiles with 
  other broad line features to probe the BLR structure 
  and kinematics. 

This paper is structured as follows: 
Section~\ref{data} describes the observations and data 
  reduction. 
Section~\ref{analysis} presents the methodology adopted 
  to convolve the NIR \feii\ template and its 
  application to the galaxy sample and results. 
Section~\ref{excitation} discusses the excitation mechanisms 
  of the NIR \feii\ emission. 
Section~\ref{location} discusses the kinematics of 
  the BLR based on the \feii\ lines and other BLR 
  emission as well the distance of the \feii\ emitting line region.
Conclusions are given in Section~\ref{conclusion}.

\section{Observations and data reduction} \label{data}
The 25 AGNs that compose our sample were selected primarily 
  from the list of \cite{jol91}, who collected data for 
  about 200 AGNs (Seyfert\,1 and quasars) to study the 
  relationship between \feii\ and radio emission. 
Additional selection criteria were applied to that initial
  sample such as the targets have to be brighter than K=12 
  mag to obtain a good compromise between signal-to-noise 
  (S/N) and exposure time for  
  the NASA 3\,m Infrared Telescope Facility (IRTF) atop 
  Mauna Kea. 
We also applied the restriction that the FWHM of the broad 
  \hb\ component of the galaxies be smaller than 
  3000\,km\,s$^{-1}$ in order to avoid severe blending of 
  the \feii\ lines with adjacent permitted and 
  forbidden features. 
Because of the last criterion, our final sample was 
  naturally dominated by narrow-line Seyfert\,1 galaxies.  
Basic information for the galaxy sample is listed in 
  Table~\ref{tab:basicdata}.

The NIR observations and data reduction for the above sample 
  will be described first. 
Non-contemporaneous optical and UV spectroscopy 
  obtained mostly from archival data (as well as pointed 
  observations) were also collected for most sources of 
  the sample for the purpose of assessing the optical and 
  UV \feii\ emission.
Since the data reduction of these latter data were 
  described elsewhere we will discuss here only their 
  sources and any particular issue we found 
  interesting to mention.
  
\subsection{Near-Infrared data}
NIR spectra were obtained at IRTF from April 2002 to June 2004. 
The SpeX spectrograph \citep{ray03}, was used in the 
  short cross-dispersed mode (SXD, 0.8-2.4\,\mum). 
In all cases, the detector employed consisted of a 
  1024$\times$1024 ALADDIN 3 InSb array with a spatial 
  scale of 0.15$\arcsec$/pixel. 
A 0.8$\times$15 slit was used giving a spectral 
  resolution of 360\,km\,s$^{-1}$.
This value was determined both from the arc lamp spectra 
  and the sky line spectra and was found to be constant 
  with wavelength within 3\%. 
During the different nights, the seeing varied between 
  0.7-1$\arcsec$.
Observations were done nodding in an ABBA source pattern 
  along the slit with typical integration times from 
  120\,s to 180\,s per frame and total on-source 
  integration times between 35 and 50\,min.
Right before/after the science frames, an A0V star was 
  observed near each target to provide a telluric standard 
  at similar airmass. 
It was also used to flux calibrate the corresponding 
  object.
 
The spectral reduction, extraction and wavelength 
  calibration procedures were performed using SPEXTOOL, 
  the in-house software developed and provided by the 
  SpeX team for the IRTF community \citep{cus04}. 
1-D spectra were extracted using an aperture window of 
  0.8$\arcsec$ in size, centered at the peak of the 
  light distribution. 
Because all objects of the sample are characterized 
  by a bright central nucleus compared with the galaxy disk,
  the light profile along the spatial direction was 
  essentially point-like.
Under this assumption, SPEXTOOL corrects for small 
  shifts due to atmospheric diffraction in the position 
  of the light peak along the different orders.

The extracted galaxy spectra were then corrected for 
  telluric absorption and flux calibrated using Xtellcor 
  \citep{vac03}, another in-house software developed 
  by the IRTF team. 
Finally, the different orders of each science spectrum 
  were merged to form a single 1-D frame. 
It was later corrected for redshift, determined from the 
  average $z$ measured from the positions of 
  [S\,{\sc iii}]\,9531\,\AA{}, Pa$\delta$, 
  He\,{\sc i}\,10830\,\AA{}, \pab\ and Br$\gamma$.
Galactic extinction corrections, as determined from the
  COBE/IRAS infrared maps of \cite{sch98}, were applied 
  for each target. 
The value of the Galactic E(B-V) used for each 
  galaxy is listed in Col. 6 of Table \ref{tab:basicdata}.

\subsection{Optical and Ultraviolet data} 
Optical spectroscopy for a sub-sample of objects 
  were obtained from different telescopes, including 
  archival data from SDSS and HST, as well as our 
  own observations. 
Column 2 of Table~\ref{tab:otherdata} lists the source of 
  the optical spectroscopy. 
The purpose of this spectra is to determine the integrated 
  flux of the \feii\ blend centered at 
  4570~\AA\ and \hb\ as well as R$_{\rm 4570}$, 
  the flux ratio \feii\,$\lambda$4570/\hb.
  
In addition, UV spectroscopy for a sub-sample of sources 
  taken with HST is also employed to compare the emission 
  line profiles of \feii\ and other BLR features, 
  including some high-ionization permitted lines.

\begin{figure}
\epsscale{0.85}
\plotone{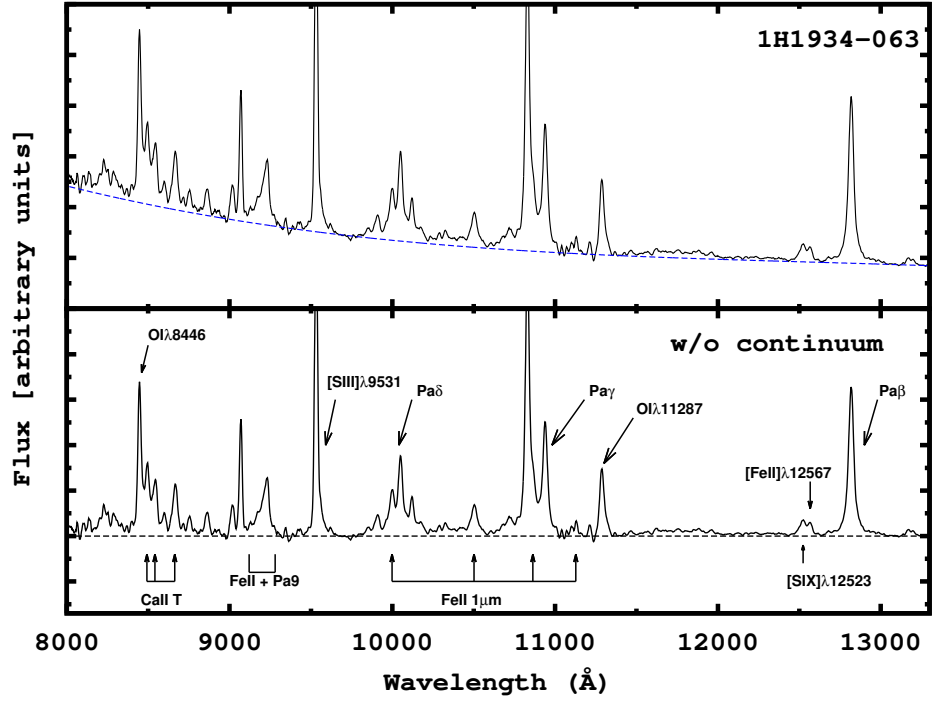}
\caption{Example of continuum subtraction and the most relevant emission lines used in this work. 
Top Panel: Observed Spectrum of 1H\,1934-063 (in the rest frame) and the continuum fit (blue dashed line). 
Bottom Panel: Spectrum of 1H\,1934-063 without the continuum. The black arrows point to the most relevant lines for this paper. 
The dashed black line indicates the zero continuum level.}
\label{fig:contsub}
\end{figure}

As our interest is in the emission line spectrum, it is 
  necessary to remove the strong continuum emission in the 
  optical and NIR, assumed to be primarily from the 
  central engine. 
To this purpose, we fit a polynomial function to the 
  observed continuum using as anchor points 
  regions free of emission lines and 
  subtract it from the spectrum. 
Figure \ref{fig:contsub} shows an example of this 
  procedure applied to 1H\,1934-063A. 
This procedure proved to be successful to our purposes. 
The analysis of the individual continuum components 
  (i.e., AGN, dust and stellar population) are 
  beyond the scope of this paper, so no effort was 
  made at interpreting the physical meaning of the fit.

In none of the cases the NIR and optical/UV spectroscopy 
  were contemporaneous. 
Therefore, no effort was made to match the continuum emission 
  in the overlapping regions of the spectra 
  (i.e., NIR and optical and UV and optical) because of 
  variability, seeing and aperture effects. 
However, since the optical data is used to provide 
  quantities to be compared to the NIR, it is important to 
  consider variability effects on the emission lines that 
  are being analyzed.
Few works in the literature, though, have found optical 
  \feii\ variability, and the overall statistics are 
  scarce.

\cite{gist96}, for example, reported variations of the 
  \feii\ bump at 5200\,\AA{} of less than 15$\%$.
\cite{die93} detected no significant variations in the 
  \feii\ lines in NGC\,5548.
\cite{biko99} found that optical \feii\ lines 
  remained constant over a 10 year monitoring campaign, 
  even when the Balmer lines and continuum were seen to 
  vary over a range of 2 to 5.
\cite{wan05} found that the \feii\ 
  variations in NGC\,4051 correlate with variations 
  in the continuum and the \hb\ line.
Similar results were found by \cite{sha12} for Ark\,564.
  
Few AGNs show strong \feii\ 
  variations (larger than 50$\%$), particularly in very 
  broad line objects 
  \citep[FWHM of H$\beta>4000$\,km\,s$^{-1}$;][]{kol81,kofr85,kol00}.
For example, \cite{kue08} carried out a reverberation 
  analysis of Ark\,120.
They were unable to measure a clean reverberation 
  lag for this object.
\cite{bar13}, though, detected \feii\ reverberation for two broad line AGNs, 
  Mrk\,1511 and NGC\,4593, using data from the \textit{LICK 
  AGN Monitoring Project}. They found variability with an amplitude lower than 20$\%$ relative to the 
  mean flux value. 
In addition, \cite{hu15} report the detection of significant \feii\ time lags
  for a sample of 9 NLS1 galaxies in order to study AGNs with high 
  accretion rates.
Difficulties in detect variations 
  in the \feii\ bump at $\lambda$4570 
  are usually ascribed to residual variations of the 
  He\,{\sc ii}\,$\lambda$4686, which is blended with \feii\ in this wavelength 
  interval \citep{kodi96}.

The general scenario that emerges from these works 
  is that the amplitude of the \feii\ variability 
  (when it varies) in response to continuum variations is 
  much smaller than that of \hb. Indeed, this 
  latter line is widely known to respond to continuum 
  variability \citep{kas00,pet04,pet98,kol00,kol06}. 
\cite{kol06} analyzed this effect using a sample of 45 
  AGNs in order to study the BLR structure.
Considering the mean values of all fractional 
  variabilities presented in their work as 
  representative of the variability effect in our 
  sample, we estimate an uncertainty of $\sim$11$\%$ 
  on the optical fluxes due to variability.
This value is also in good agreement with the 
  results of \cite{hu15}, which found an average 
  fractional variability of 10$\%$ for \feii\
  when compared with \hb.
This value is within the error in the line 
  fluxes measured in this work; therefore, we conclude 
  that variability is unlike to impact our results.

\section{Analysis Procedure} \label{analysis}
\subsection{NIR \feii\ Template Fitting} \label{tempfit}
Modeling the \feii\ pseudo-continuum, formed by the 
  blending of the thousands of \feii\ multiplets, 
  remains a challenge for the analysis of this emission 
  since it was first observed by \cite{waok67}. 
\cite{sar68} noted that I\,Zw\,1, for instance, had the 
  same kind of emission but with stronger and narrower 
  lines.
The strength of the \feii\ lines in that AGN mades it a 
  prototype of the strong \feii\ emitters as well 
  as of the NLS1 subclass of AGNs, leading to the 
  development of empirical templates of this emission based 
  on this source \citep{bor92,ves01,ver04,gar12}.

\cite{sig03} and \cite{sig04} presented the first \feii\ model from the UV 
  to the NIR using an iron atom with 827 fine structure energy levels 
  and including all known excitation mechanisms 
  (continuum fluorescence via the UV resonance lines, 
  self-fluorescence via overlapping \feii\ transitions, and 
  collisional excitation) that were traditionally considered by \cite{wil85} and \cite{bal04} 
   in addition to fluorescent excitation by \lya\ as suggested by \cite{pen87}.
Their models incorporates photoionization cross sections 
  \citep[references in][]{sig04} that include a large number of 
  autoionizing resonances, usually treated as lines but too numerous 
  to count explicitly, which means that they include many more 
  photo-excitation transitions than the 23,000 bound-bound lines and 
  would form part of the pseudo-continuum.
Moreover, in their work they show how the \feii\ intensity
  varies as a function of the ionization parameter ($U_{\rm ion}$)\footnote{The ionization parameter $U_{\rm ion}$ is correlated with $\Phi_H$, the flux of 
  hydrogen ionizing photons at the illuminated face of the cloud, by $U_{ion}=\Phi_H/n_Hc$.}, 
  the particle density ($n_H$) and the microturbulence velociy ($\zeta_t$).
  Details of all the physics involved in the calculations are in \cite{sig03}.
\cite{lan08} were the first to confront these models with 
  observations, noting some discrepancies between the model 
  and the observed emission lines. 
\cite{bru08} using an \feii\ model with 830 energy levels (up to 14.06\,eV)
  and 344,035 atomic transitions, found that the model parameters that best fit the 
  observed UV spectrum of I\,Zw\,1 were
  $log(\Phi_H)=20.5$, $log(n_H)=11.0$~cm$^{-3}$ and $\zeta_t$=20\,km\,s$^{-1}$.
\cite{gar12} modeled the observed NIR spectrum of I\,Zw\,1 using 
  a grid of Sigut $\&$ Pradhan's templates covering several 
  values of ionization parameters and particle 
  densities, keeping the microturbulence velocity constant at $\zeta_t$=10\,km\,s$^{-1}$.
They found that the model with $U_{\rm ion}$=$-2.0$ (implying in $log(\Phi_H)=21.1$) 
  and $log(n_H)$=$12.6$ cm$^{-3}$ best fit the observations.
Note that these values are comparable to those found by \cite{bru08}, suggesting
that the physical conditions of the clouds emitting \feii\ are similar.

The template developed by \cite{gar12} is composed of 
  1529 \feii\ emission lines in the region between 
  8300\,\AA{} and 11600\,\AA{}.
In order to apply it to other AGNs, it is first necessary 
  to convolve it with a line profile that is representative 
  of the \feii\ emission.
At this point we are only interested in obtaining a mathematical 
  representation of the empirical profiles.
In order to determine the optimal line width, 
  we measured the FWHM of individual iron lines detected 
  in that spectral region. 
We assumed that each \feii\ line could 
  be represented by a single or a sum of individual 
  profiles and that the main source of line broadening was the Doppler 
  effect.
  
The \feii\ emission lines $\lambda$10502 and 
  $\lambda$11127 are usually isolated and allow an accurate 
  characterization of their form and width. 
The LINER routine \citep{pog93}, a $\chi^2$ minimization 
  algorithm that fits up to eight individual profiles 
  (Gaussians, Lorentzians, 
  or a linear combination of them as a pseudo-Voigt profile) 
  to a single or a blend of several line profiles, was used 
  in this step.
We found that a single Gaussian/Lorentzian profile was enough to 
  fit the two lines above in all objects in the sample. 
However, the difference between the \textit{rms} error for 
  the Gaussian and Lorentzian fit was less than 5\%, 
  which lies within the uncertainties. 
As \feii\,$\lambda$11127 is located in a spectral 
  region with telluric absorptions, residuals left after 
  division by the telluric star may hamper 
  the characterization of that line profile. 
For this reason, we considered \feii\,$\lambda$10502 
  as the best representation of the broad \feii\ 
  emission. 
Note that \cite{gar12} argued that the \feii\ 
  $\lambda$11127 is a better choice than 
  \feii\,$\lambda$10502 because the latter can be 
  slightly broadened due to a satellite \feii\
  line at $\lambda$10490. 
However, \feii\,$\lambda$10490 is at least 5 
  times weaker relative to $\lambda$10502 
  \citep{gar12,rod02}. 
Therefore, we adopted the $\lambda$10502 line as 
  representative of the \feii\ profile because it 
  can be easily isolated and displays a good S/N in 
  the entire sample. 
The flux and FWHM measured for this line  
  are shown in Columns 2 and 3 of Table \ref{tab:linefit}.
  
For each object, a synthetic \feii\ spectrum was 
  created from the template using the FWHM listed in 
  Table~\ref{tab:linefit} and then scaled to the 
  integrated line flux measured for the $\lambda$10502 
  line.
 
In order to ensure that the line width used to convolve 
  the template best represented the FWHM of the 
  \feii\ emission region, we generated for each galaxy 
  a grid of 100 synthetic spectra with small variations in 
  the line width (up to 10\% around the best value) for 3 
  different functions (Gaussian, Lorentzian and Voigt). 
In all cases, the value of the FWHM ​​that minimized the 
  \textit{rms} error after subtraction of the template was very 
  close (less than 1\%) to the width found from the direct 
  measurement of the $\lambda$10502 line. 
Also, the best profile function found in all cases was the same 
  one that fitted initially. 
The final parameters of the convolution of the template for 
  each source are shown in Table~\ref{tab:temppar}.

Figures~\ref{fig:nirtemp1} to \ref{fig:nirtemp4} show the 
  observed spectra and the template convolved with the 
  best parameters (upper panel). 
The ion-free spectra after subtraction of the modeled 
  \feii\ emission are shown in the bottom panels. 
It can be seen that overall the template nicely reproduces 
  the observed \feii\ in all AGNs.
We measure the \textit{rms} error of the subtraction of the 
  template using as reference the regions around the 
  1~$\mu$m lines.
The mean \textit{rms} error of the template subtraction are shown 
  in column 4 of Table~\ref{tab:temppar}.

From the best matching template we estimated the flux of 
  the 1\,\mum\ lines. 
Columns 2-6 of Table~\ref{tab:1mulines} show the fluxes 
  of each line. 
We define the quantity \Rone\ as the ratio between the 
  integrated flux of the 1\,\mum\ lines and the flux of the 
  broad component of \pab. 
This value is presented in column 7 of 
  Table~\ref{tab:1mulines}. 
We consider this ratio as an indicator of the NIR 
  \feii\ strength in each object.
Sources with weak \feii\ emission are characterized 
  by low values of \Rone\ (0.1 $-$ 0.9) while strong 
  \feii\ emitters display values of R$_{1\mu m} >$1.0.
  
Two features in the residual spectra (after subtraction of  
  the \feii\ emission) deserve comments.
The first one is the apparent \feii\ excess centered 
  at 11400\,\AA{} that is detected in some sources.
We identify this emission with an \feii\ line 
  because it was first identified in I\,Zw\,1 
  but it is absent in sources with small \Rfour.
Therefore, its detection may be taken as an indication of a 
  I\,Zw\,1-like source.   
The 11400\,\AA{} feature is formed by a blend of 
  8 \feii\ lines, being those located at $\lambda$11381.0 and 
  $\lambda$11402.0 the strongest ones. They both carry 95$\%$ of the
  predicted flux of this excess. 
\cite{gar12} had to modified it to properly reproduce the observed strength in 
  that object because the best matching \feii\ model 
  severely underestimated it.
Nonetheless, when the template was applied to Ark\,564, the 
  feature was overestimated.
Our results show that the peak at 11400\,\AA{} is present 
  only in the following objects: Mrk\,478, PG\,1126-041, 
  PG\,1448+273, Mrk\,493 and Ark\,564.
In the remainder of the sample it is absent.

The second feature is the \feii\ bump centered at $\lambda$9200, 
  which is actually a blend of aproximately 80 \feii\ lines plus Pa9.
The most representative \feii\ transitions in this region are 
  \feii$\lambda$9132.36, $\lambda$9155.77, $\lambda$9171.62, $\lambda$9175.87, $\lambda$9178.09, 
  $\lambda$9179.47, $\lambda$9187.16, $\lambda$9196.90, $\lambda$9218.25 and $\lambda$9251.72.
In order to assess the suitability of the template to 
  reproduce this feature, we modeled the residual left in 
  the 9200\,\AA{} region after subtracting the template. 
The residual was constrained to have the same profile and 
  FWHM as \pab. 
The flux of Pa9 line should be, within the uncertainties, the flux 
  of \pab\ multiplied by the Paschen decrement factor 
  (\pab/Pa9\,$\sim$\,5.6) \citep{gar12}.
The results obtained for each object are shown in column 3 of 
  Table~\ref{tab:9200fit}. 
  
Column 4 of Table~\ref{tab:9200fit} shows the expected flux 
  for that line. 
When we compare it to the measured flux, we find that the 
  latter is systematically larger, which indicates 
  that the template underestimates the value of 
  the \feii\ emission in this region.
Similar behavior was observed by \cite{mar15} for a 
  a smaller spectral region (0.8-0.95\,$\mu$m).
They subtracted the NIR \feii\ emission in 14 luminous AGNs 
  using the template of \cite{gar12} for this region and 
  found an excess of \feii\ in the 9200\,\AA{} bump after 
  the subtraction.

Nevertheless, we can estimate the total \feii\ emission 
  contained in the 9200\,\AA{} bump using the residual flux left 
  after subtraction of the expected Pa9 flux 
  to ``top up'' the flux found in the \feii\ template.
Column 5 of Table~\ref{tab:9200fit} lists this total \feii\ 
  flux in the bump. 
We define the quantity \Rnine\ as the flux ratio between the 
  \feii\ bump and the broad component of \pab. 
The results are listed in column 6 of Table~\ref{tab:9200fit}.
  
Except for the residuals in the $\lambda$9200 region, 
  our results demonstrate the suitability of the semi-empirical 
  NIR \feii\ template in reproducing this emission in a 
  large sample of local AGNs. 
The only difference from source to source are scales factors 
  in FWHM and flux, meaning that the relative intensity 
  between the different \feii\ lines remains approximately 
  constant, similar to what is observed in the UV and optical 
  region \citep{bor92,ves01,ver04}. 
Figures~\ref{fig:nirtemp1}-\ref{fig:nirtemp4} also confirm that 
  the template developed for the \feii\ emission in the 
  NIR can be applied to a broad range of Type~1 objects. 
  
The results obtained after fitting the \feii\ template 
  allow us to conclude that: 
({\it i}) without a proper modeling and subtraction of that 
  emission, the flux and profile characteristics of other adjacent BLR 
  features can be overestimated; 
({\it ii}) once a good match between the semi-empirical 
  \feii\ template and the observed spectrum is found, 
  individual \feii\ lines, as well as the NIR \feii\ 
  fluxes, can be reliably measured; 
({\it iii}) the fact that the template reproduces well the 
  observed NIR \feii\ emission in a broad range of AGNs 
  points to a common excitation mechanisms for 
  the NIR \feii\ emission in Type\,1 sources.

\subsection{Emission line fluxes of the BLR in the NIR} \label{profiles}

Modelling the pseudo-continuum formed by 
  numerous permitted \feii\ lines in the 
  optical and UV regions is one of the most challenging 
  issues for a reliable study of the BLR. 
Broad optical emission lines of ions other than 
  \feii\ are usually heavily blended with \feii\ 
  multiplets and NLR lines. 
In order to measure their fluxes and characterize 
  their line profiles, a careful removal of the \feii\ 
  emission needs to be done first. 
In this context, the NIR looks more promising for the 
  analysis of the BLR at least for three reasons. 
First, the same set of ions detected in the optical are 
  also present in that region (H\,{\sc i}, He\,{\sc i}, 
  \feii, He\,{\sc ii} in addition to \oi\ and \caii, not seen 
  in the optical).
Second, the lines are either isolated or moderately blended 
  with other species.
Third, the placement of the continuum is less prone
  to uncertainties relative to the optical because the 
  pseudo-continuum produced by the \feii\ is weaker. 

This section will describe
  the method employed to derive the flux of the most 
  important BLR lines in the NIR after the removal of 
  all the emission attributed to \feii.  

To this purpose, the presence of any NLR emission should be 
  evaluated first, and, if present, subtracted from the 
  observed lines profiles.
An inspection of the spectra reveals the presence of 
  forbidden emission lines of [S\,{\sc iii}]\,$\lambda$9068 
  and $\lambda$9531 in all sources analyzed here. 
Therefore, a narrow component is also expected for the 
  Hydrogen lines that may contribute a non-negligible 
  fraction to the observed flux. 
In order to measure this narrow component, we followed the 
  approach of \cite{rod00}, which consists of adopting the 
  observed profile of an isolated NLR line as a template, 
  scaling it in strength, and subtracting it from each 
  permitted line. 

Note that neither \oi, \caii, nor \feii\ 
  required the presence of a narrow component to model their 
  observed profiles, even in the spectra with the best S/N. 
In all cases, after the inclusion and subtraction of a 
  narrow profile, an absorption dip was visible in the 
  residuals. 
\cite{rod02} had already pointed out that no contribution 
  from the NLR to these lines is expected as high gas densities 
  ($>$10$^{8}$ cm$^{-3}$) are necessary to drive this emission.
This result contrasts to claims made by \cite{don10}, who 
  include a narrow component, with origin in the NLR, in the 
  modelling of the optical \feii\ lines.
   
For consistency, \oi\,$\lambda$8446 
\footnote{Note that this line is actually a closely spaced triplet of \oi\,$\lambda$8446.25, $\lambda$8446.36 and $\lambda$8446.38)}
  and \caii\,$\lambda$8498,$\lambda$8542 had their widths 
  (in velocity space) constrained to that of  
  \oi\,$\lambda$11287 and  
  \caii\,$\lambda$8662, respectively. 
These latter lines are usually isolated and display 
  good S/N. 
Note however that for a part of our sample, it was not 
  possible to obtain a good simultaneous fit to the 
  three calcium lines using this approach. 
We attribute this mostly to the fact that some of the AGNs 
  have the \caii\ lines in absorption \citep{pen87}
  and also to the lower S/N of 
  \caii\,$\lambda$8498,$\lambda$8542 
  as they are located in regions with reduced atmospheric 
  transmission. 

\begin{figure}
\epsscale{0.85}
\plotone{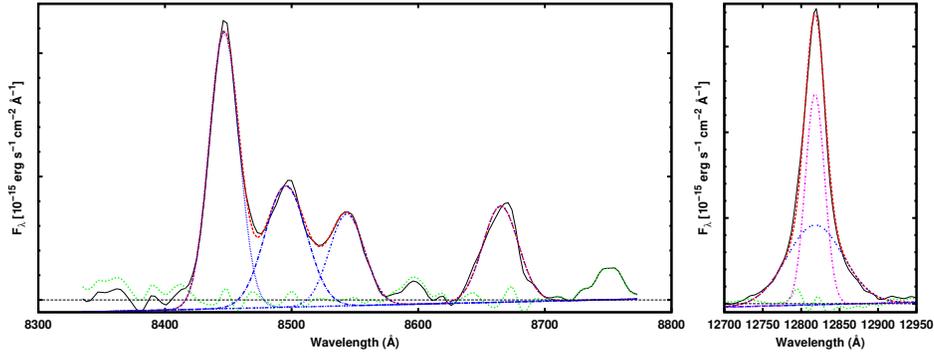}
\caption{Example of line deblending for 1H\,1934-063. 
The left panel shows each line of the \caii\ triplet 
(blue dotted-dashed line), the \oi\,$\lambda$8446 line 
(blue dotted line) and the total fit of these lines 
(red dashed line).
Right panel shows the deblend of the \pab\ line in broad 
(two-dotted blue line) and narrow (dotted-dashed magenta line) 
components. The red dashed line is the sum of these two 
components.}
\label{fig:deblend}
\end{figure}

Table \ref{tab:linefit} lists the fluxes and FWHM 
  measured for the most conspicuous emission lines of our sample. 
The errors presented are due to the small variations to stablish the continuum zero 
  level for the fits. 
Figure \ref{fig:deblend} shows an example of the deblending procedure 
  applied to each of the lines analyzed.

\section{\feii\ Excitation Mechanism: 
	    \lya\ fluorescence 
	    and Collisional Excitation} \label{excitation}

\begin{figure}
\epsscale{0.85}
\plotone{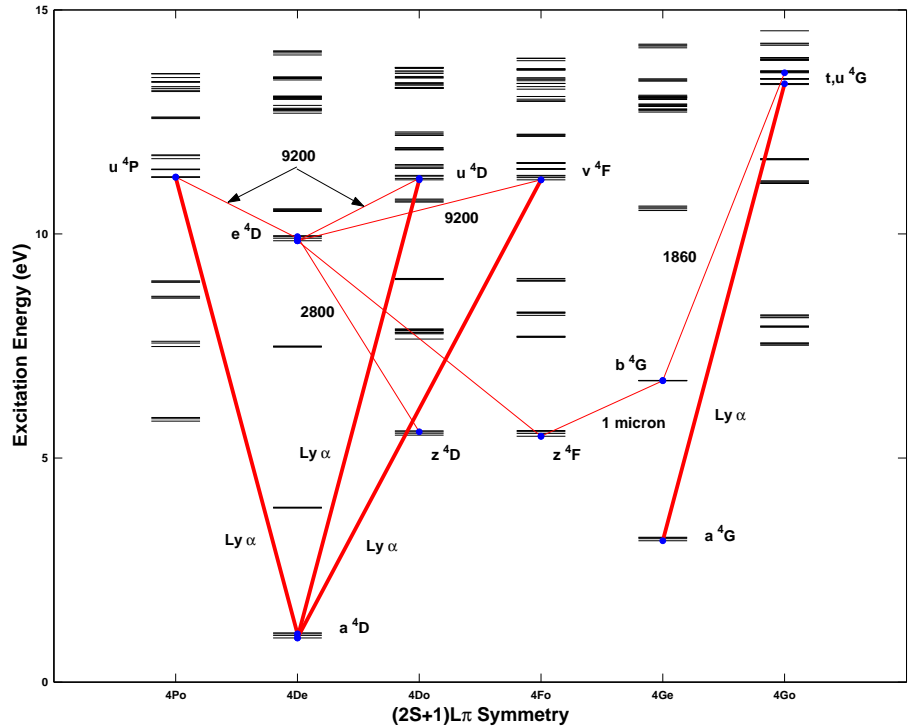}
\caption{Partial Grotrian diagram for the \feii\ system showing the 
transitions forming the 1\,\mum\ lines and the bump at 9200\,\AA{}.
The \lya\ fluorescence route and the subsequent cascades. 
The 1\,\mum\ lines are produced by the multiplets transitions 
b$^4$G$\rightarrow$z($^4$F, $^4$D). The \feii\,$\lambda$9200 bump is produced by the transitions 
u($^4$P,$^4$D),v$^4$F$\rightarrow$e$^4$D. Figure from \cite{sig03}}
\label{fig:gothrian}
\end{figure}

The primary excitation mechanism invoked to explain most 
  of the NIR \feii\ lines is \lya\ fluorescence 
  \citep{sipr98,sig03,rod02}. 
In this scenario the iron lines are produced by primary
  and/or secondary cascading after the absorption of a 
  \lya\ photon between the levels 
  a$^4$G $\rightarrow$ (t,u)$^4$G$^0$ 
  and a$^4$D $\rightarrow$ u$^4$(P,D),v$^4$F.
As can be seen in Figure~\ref{fig:gothrian}, 
  there are two main NIR \feii\ features in the range of 
  0.8-2.5\,$\mu$m that arise from this process: the 
  1\,\mum\ lines and the bump centered in $\lambda$9200.
The importance of this excitation channel is the fact 
  that it populates the upper energy levels whose 
  decay produces the optical \feii\ lines, traditionally used 
  to measure the intensity of the iron emission in AGNs
  \citep{sig03}. 
Much of the challenge to the theory of the \feii\ emission 
  is to determine if this excitation channel is indeed 
  valid for all AGNs and the degree to which this process 
  contributes to the observed \feii\ flux.

In order to answer these two questions, we will analyze first 
   the 1$\mu$m lines.
They result from secondary cascading after the capture of a 
  \lya\ photon that excites the levels 
  a$^4$G $\rightarrow$ (t,u)$^4$G$^0$ 
  followed by downward UV transitions to the level b$^4$G via 
  1870/1873\,\AA{} and 1841/1845\,\AA{} emission and finally 
  b$^4$G $\rightarrow$ z$^4$F$^0$ transitions, which produce 
  the 1\,\mum\ lines. 
These lines are important for at least two reasons: 
(\textit{i}) they are the most intense NIR \feii\ lines that 
  can be isolated; and
(\textit{ii}) after they are emitted, the z$^4$F and z$^4$D 
  levels are populated. 
  These levels are responsible for $\sim$50$\%$ of the total 
  optical \feii\ emission.
Therefore, the comparison between the optical and NIR 
  \feii\ emission can provide important clues about 
  the relevance of the \lya\ fluorescence process in 
  the production of optical \feii. 

\begin{figure}
\epsscale{0.85}
\plotone{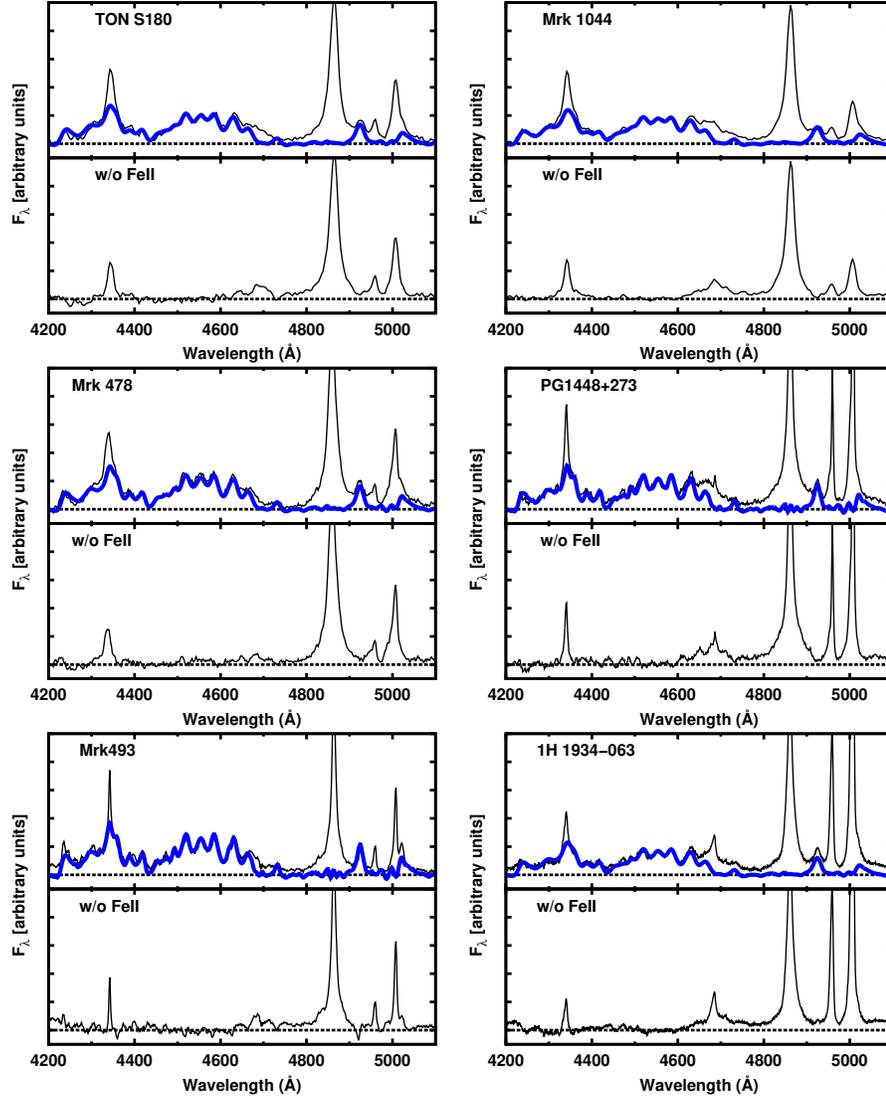}
\caption{Examples of The convolved optical \feii\ template. Top panels show the continuum-subtracted 
spectrum of the AGN (black line), with the optical spectrum of \feii\ (blue bold line) superposed. 
Bottom panels show the spectrum of each object 
without this emission.}
\label{fig:opttemp}
\end{figure}

We measured the optical \feii\ emission for 18 out of 25 AGNs 
  in our sample by applying the \citet{bor92} 
  method to the optical spectra presented in 
  Section~\ref{data}.
\cite{bor92} found that a suitable \feii\ template 
  can be generated by simply broadening the \feii\ 
  spectrum derived from the observations of I\,Zw\,1. 
The \feii\ template strength  
  is free to vary, but it is broadened to be consistent 
  with the width found for the NIR iron lines. 
The best \feii\ template is found by minimization 
  of the $\chi^2$ values of the fit region, set
  to 4435$-$4750\,\AA. 
Half of the lines that form 
  this bump comes from downward cascades from the z$^4$F levels. 
Figure~\ref{fig:opttemp} shows an example of the optical 
  \feii\ template fit to the observed spectrum.
In addition, we measured the integrated flux of the 
  \hb\ line after subtraction of the underlying 
  \feii\ emission.
  
Afterwards, the amount of \feii\ present in 
  each source was quantified by means of \Rfour, 
  the flux ratio between the \feii\ blend centered 
  at 4570~\AA\ and \hb. 
Currently, this quantity is employed as an estimator of
  the amount of optical iron emission in active galaxies. 
Although values of \Rfour\ for some objects of our 
  sample are found in the literature \citep{jol91,bor92}
  we opted for estimating it from our own data. 
Differences between values of \Rfour\ found for the
  same source by different authors, the lack of 
  proper error estimates in some of them, and the use of 
  different methods to determine \Rfour\ 
  \citep{per88,jol91}, encouraged us to this approach. 
The values found for \Rfour\ in our sample are listed 
  in Table~\ref{tab:4570fit}.

Model results of \cite{bru08} show that both the BLR and the NLR 
  contribute to the observed permitted Fe\,{\sc ii} emission in the 4300-5400~\AA\
  interval. The contribution of the NLR is particularly strong in the 4300-4500\AA{} region and would 
  arise from the a($^6S,^4G$)$\rightarrow$a($^6D,^4F$) and b$^4F\rightarrow$a$^6D$ transitions, in 
  regions of low density ($n_{\rm H} < 10^{4}$ cm$^{-3}$) gas. The iron BLR component, in contrast, dominates 
  the wavelength interval 4500-4700\AA{}. This hypothesis was
  tested by \cite{bru08} in the NLS1 galaxy I\,Zw\,1. 	
Our optical spectra includes the interval of 4200-4750\,\AA{}, and 
   recall that we followed the empirical method proposed by \citet{bor92}
   to quantify this emission.
   In other words, no effort was made to separate the BLR and NLR components.
  Moreover, because the relevant 
  optical quantity in our work is composed by the blends of \feii\ lines
  located in the interval 4435-4750~\AA,
  were the NLR almost do not contribute,
  we conclude that this NLR component, if it exists, does not interfer 
  in our results.
It is possible, however, to test the presence of NLR \feii\ emission in the
  NIR. \cite{rif06}, for instance, in their NIR atlas of 52 AGNs clearly identified the forbidden 
  [Fe\,{\sc ii}] lines at 12570\,\AA{} and 16634\,\AA{} in most objects of their sample, but did not find 
  evidence of permitted \feii\ emission from the NLR. 
  Here, we also confirm this result. For none of the NIR 
  spectra studied here evidence of a narrow component was found, even in isolated 
  \feii\ lines such as \feii$\lambda$10502.
  If this contribution exists, it should be at
  flux levels smaller than our S/N.
  
\begin{figure}
\epsscale{0.85}
\plotone{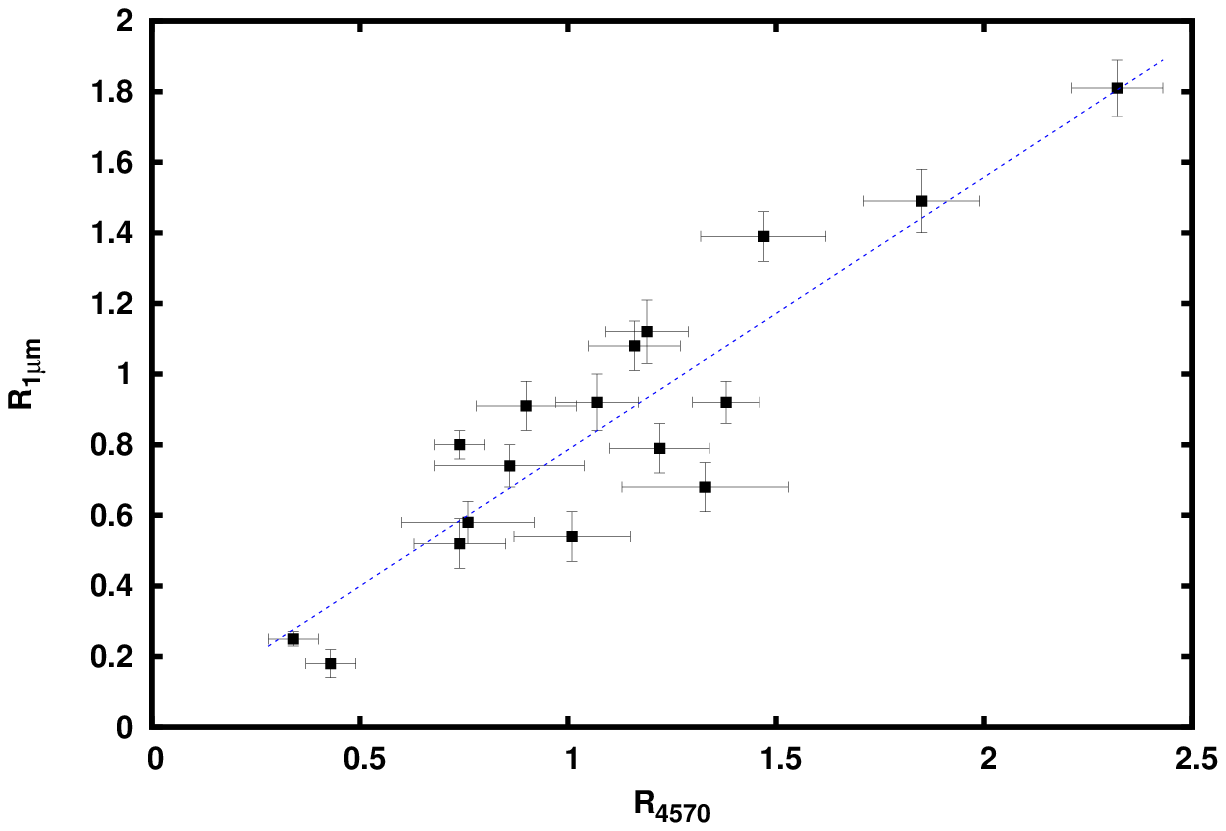}
\caption{Correlation between \Rone\ and \Rfour. The blue dotted line shows the best linear fit.}
\label{fig:RonevsR4570}
\end{figure}

Table \ref{tab:1mulines} lists the fluxes of the 
  1\,\mum\ lines. 
As in the optical, we derive the quantity \Rone. 
In order to determine if both ratios are correlated, we 
  plot \Rone\ vs \Rfour\ in Figure~\ref{fig:RonevsR4570}.
Since the energy levels involved in producing the optical lines in 
  the \Rfour\ bump are populated after the emission 
  of the NIR \feii\ 1$\mu$m lines, a correlation between 
  these quantities can be interpreted as evidence of a 
  common excitation mechanism. 
  
An inspection to Figure~\ref{fig:RonevsR4570} shows that 
  \Rone\ and \Rfour are indeed strongly correlated, 
  at least for the range of values covered by our sample.
  
In order to obtain a linear fit and determine the 
  correlation coefficient, we perform a Monte Carlo 
  simulation with the \textit{bootstrap} method 
  \citep[similarly to][]{bee90}. 
First, we run 10000 Monte Carlo simulations in order to 
  determine the effect of the \Rone\ and \Rfour\ uncertainties 
  in the linear fit. 
For each realization, random values of these two quantities 
  were generated (constrained to the error range of the 
  measurements) and a new fit was made. 
The standard deviation of the fit coefficients, 
  $\epsilon_i$, was determined and represents the 
  uncertainty of the values over the linear fit 
  coefficients.
The next step was to run the bootstrap realizations in 
  order to derive the completeness and its effects on 
  the fit.
For each run, we made a new fit for a new sample randomly 
  constructed with replacement from the combination of the 
  measured values of \Rone\ and \Rfour. 
The standard deviation of these coefficients, $\epsilon_e$, 
  gives us the intrinsic scatter of the measured values. 
Finally, the error in the coefficients are given by the 
  sum of the $\epsilon_i$ and $\epsilon_e$ in quadrature, 
  i.e., $\sqrt{\epsilon_e ^2 + \epsilon_i ^2}$. 
The strength of the correlation can be 
  measured by the Pearson Rank coefficient, 
  which indicates how similar two sample populations 
  are.
  
Following the method above, we found a Pearson rank 
  coefficient of P = 0.78 for the correlation between 
  \Rone\ and \Rfour.
This suggests that the two emissions are very likely 
  excited by the same mechanisms.
However, it does not prove that \lya\ fluorescence 
  is the dominant process.
This is because collisional excitation is also an 
  option.
\cite{rod02}, using HST/FOS spectra, found that the 
  \feii\ UV lines at 1860\,\AA\ were intrinsically weak, 
  pointing out that \lya\ fluorescence could not 
  produce all the observed intensity of the 1\,\mum\ lines 
  because the number of photons of the latter were 
  significatively larger than those in the former 
  (see Figure~\ref{fig:gothrian}). 
They concluded that collisional excitation was responsible 
  for the bulk of the \feii\ emission. 
   
We inspected the UV spectra available for our sample 
  in the region around 1860\,\AA{}. 
The evidence for the 
  presence of these lines is marginal.
For four objects in our sample, though, it was possible 
  to indentify them: 1\,H1934-063, Mrk\,335, Mrk\,1044 and Ark\,564. 
The upper limit of such UV emission in these galaxies ranges from
  0.6 to 13.2 (10$^{-14}$\,ergs\,cm$^{-2}$\,s$^{-1}$) 
  for Mrk\,1066 and 1\,H1934-063, respectively \citep{rod02}.
This does not necessarily mean that the lines are not actually 
  emitted in the remainder of the sample.
Extinction, for instance, can selectively absorb photons 
  in the UV relative to that of the NIR. 
Also, the region where these UV lines are located, at 
  least for the spectra we have available, is noisy 
  and makes any reliable detection of these lines very 
  difficult. 
  
Taking into account that the 1\,\mum\ lines are strong 
  in all objects while the primary cascade lines from 
  which they originate are marginally detected, we 
  conclude that \lya\ fluorescence does not dominate 
  the excitation channel leading to the NIR \feii\ emission. 
  
Here, we propose that collisional excitation is the main 
  process behind the iron NIR lines. 
This mechanism is more efficient at temperatures above 
  7000\,K \citep{sig03}.
Such values are easily found in photoinization 
  equilibrium clouds \citep[$\sim$10000\,K,][]{ost89}, 
  exciting the bound electrons from the ground levels to
  those where the 1\,\mum\ lines are produced. 
The constancy of the flux ratios among \feii\ NIR lines 
  found from object to object of our sample supports 
  this result. 
  
\begin{figure}
\epsscale{0.85}
\plotone{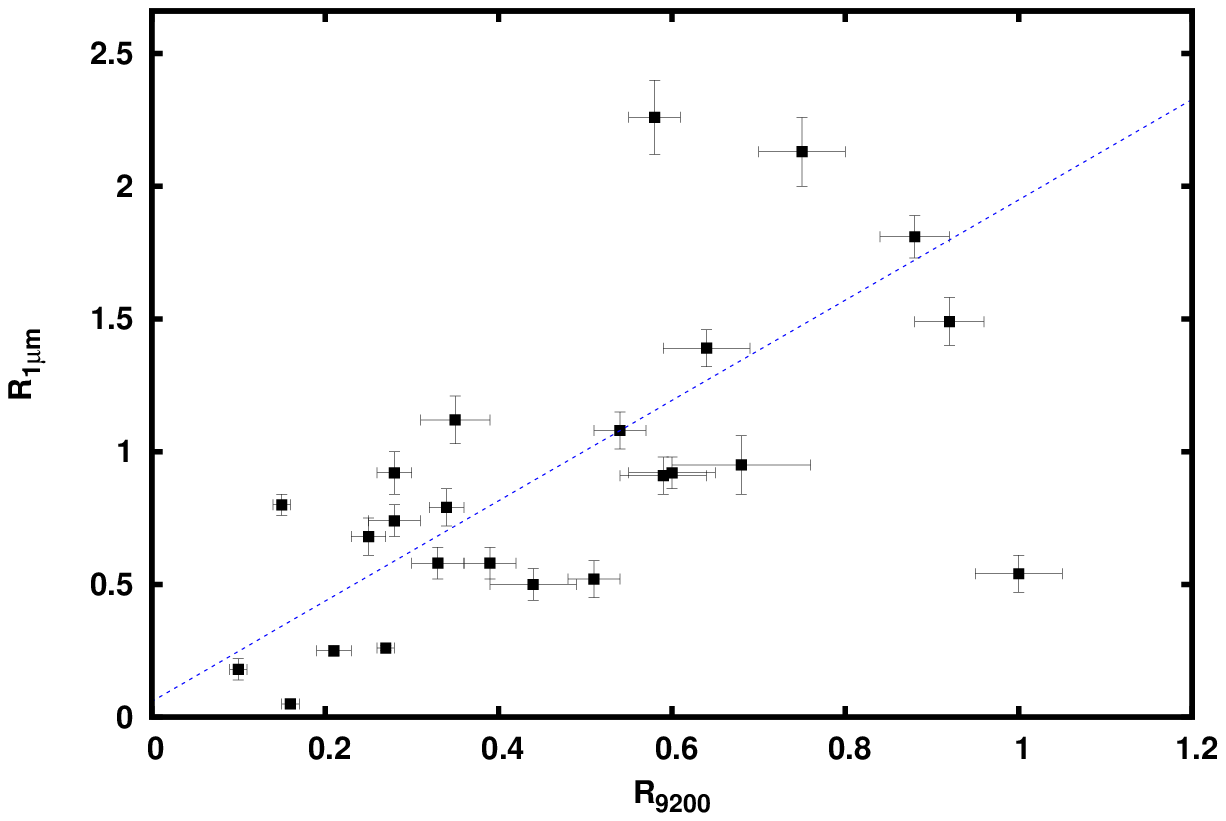}
\caption{Correlation between \Rone\ and \Rnine. The blue dotted line shows the best linear fit.}
\label{fig:RonevsR9200}
\end{figure}

\lya\ fluorescence, though, should still contribute 
  to the flux observed in the 1$\mu$m lines even if 
  it is not the dominant mechanism.
This can be observed in Figure~\ref{fig:RonevsR9200}, 
  where \Rone\ vs \Rnine\ is plot.
It can be seen that both quantities are correlated, 
  with a Pearson coefficient of P = 0.72. 
However, in order to make a crude estimate of the
  contribution of the fluorescence process, we should
  look at other relationships between the iron lines,
  such as the bumps at 9200\,\AA{} and 4570\,\AA{}.
  
\begin{figure}
\epsscale{0.85}
\plotone{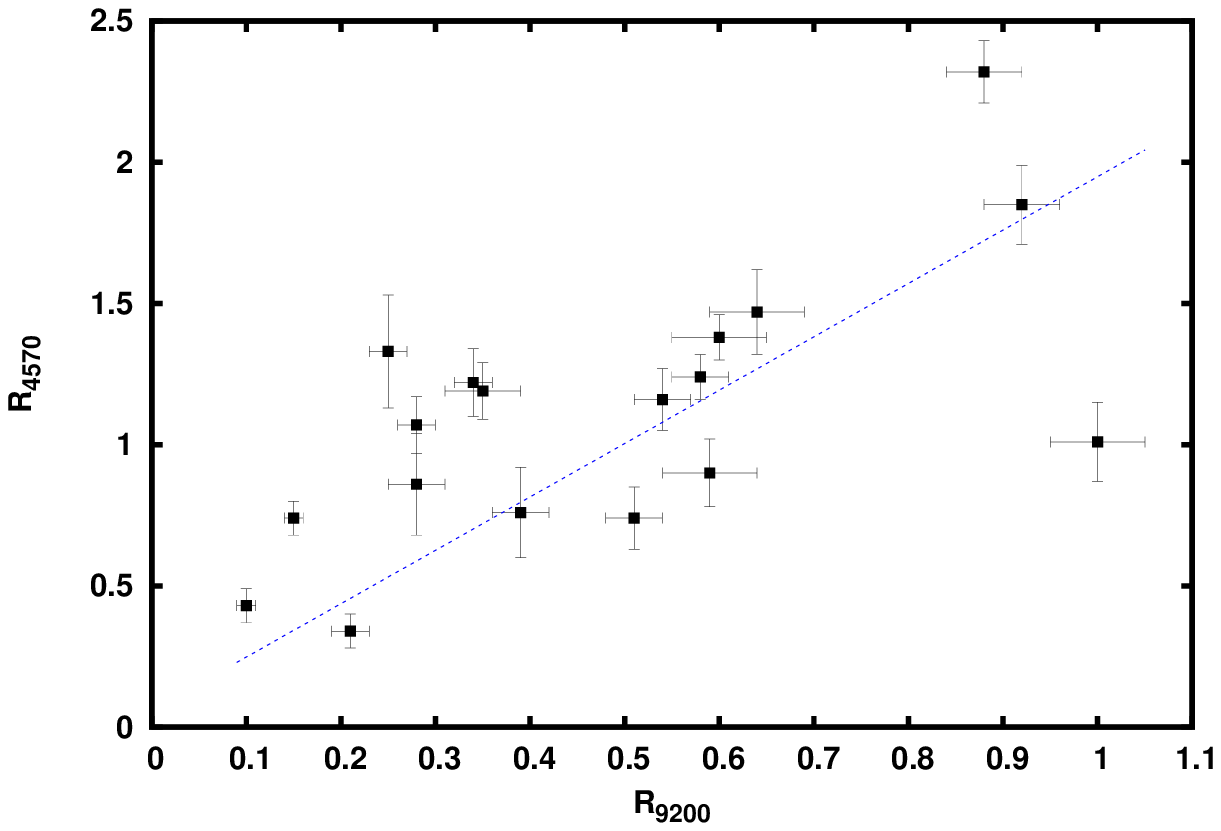}
\caption{Correlation between \Rnine\ and \Rfour. The blue dotted line shows the best linear fit.}
\label{fig:R9200vsR4570}
\end{figure}

Recall that the former is produced after the absorption 
  of a \lya\ photon, exciting  the levels 
  a$^4$D $\rightarrow$ (u,v)$^4$(D,F) followed by 
  downward transitions to the level e$^4$D via the 
  emission of the $\lambda$9200 lines. 
This latter level decays to e$^4$D $\rightarrow$ z$^4$(Z,F), 
  via UV transitions emitting the lines at $\sim$2800\,\AA{}. 
A further cascade process contributes to produce the 
  $\lambda$4570 bump. 
However, collisional excitation may also populate the upper 
  levels leading to the $\lambda$4570 bump.
As the $\lambda$9200 bump is clearly present in all objects 
  of the sample, the presence of this excitation channel is 
  demonstrated. 
In order to assess the relative contribution of the
  \lya\ fluorescence to the optical \feii\ emission,
  we plot \Rnine\ and \Rfour\ in Figure~\ref{fig:R9200vsR4570}.
It can be seen that both quantities are indeed well correlated 
  (P = 0.76), showing that part of the photons producing the
  9200\AA\ bump are converted into $\lambda$4570 photons.
  
It is then possible to make a rough estimate of the contribution of the 
  \lya\ fluorescence to the optical \feii\ emission 
  through the comparison of the number of photons observed 
  in both transitions.
Table~\ref{tab:photon} shows the number of photons in the 
  $\lambda$4570 bump (column 2) and that in the 
  $\lambda$9200 bump (column 3). 
The ratio between the two quantities is listed in column 4. 
From Table~\ref{tab:photon} we estimate that \lya\ 
  fluorescence is responsible for $\sim$36$\%$ of the 
  observed optical lines in the \feii\ bump centered at 
  4570\,\AA{}. 
The optical \feii\ bump at $\lambda$4570 represents about 
  $\sim$50$\%$ of the total optical \feii\ emission 
  \citep{ver04}. 
This means that, on average, 18$\%$ of all optical \feii\ photons 
  observed are produced via downward transitions excited by 
  \lya\ fluorescence.
This result is in agreement to that presented by 
  \cite{gar12}, which estimated a contribution of 20$\%$ 
  of this excitation mechanism in I\,Zw\,1.

\section{Location of the \feii\ 
	    Emitting Line Region} \label{location}

The fact that the BLR remains unresolved for all 
  AGNs poses a challenge to models that try to predict 
  the spatial structure and kinematics of this region. 
In the simplest approach, we assume that the 
  proximity of this region to the central source (black hole 
  plus accretion disk) implies that the movement of the gas 
  clouds is dominated by the gravitational potential of 
  the black hole. 
Under this assumption, the analysis of the line profiles 
  (form and width) can provide us with clues about the 
  physical structure of this region.

We address the above issue using the most prominent lines 
  presented in Table~\ref{tab:linefit}. 
The line profiles of \caii\,$\lambda$8664, 
  \feii\ $\lambda$10502, \oi\,$\lambda$11287 
  and \pab\ are relatively isolated or only moderately 
  blended, making the study of their line profiles more robust 
  than their counterparts in the optical. 
With the goal of obtaining clues on the structure and kinematics 
  of the BLR, we carried out an analysis of these four line 
  profiles detected in our galaxy sample. 

\begin{figure}
\epsscale{0.85}
\plotone{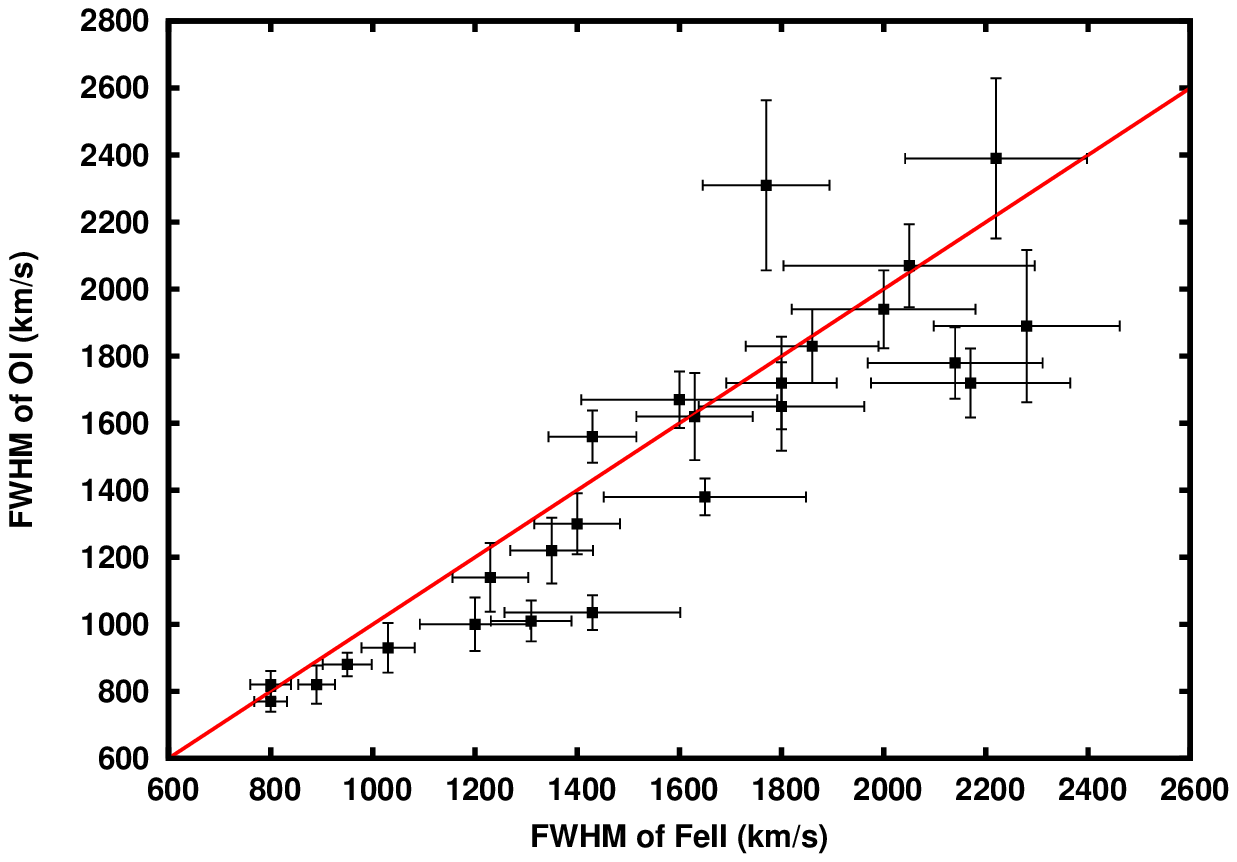}
\caption{Correlation between the FWHM of \feii\ and the FWHM of \oi. The red line shows the unitary relationship.}
\label{fig:feiivsoi}
\end{figure}

Figure~\ref{fig:feiivsoi} shows the FWHM of \oi\
  vs that of \feii. 
It is clear from the plot that both lines have very 
  similar widths, with the locus of points very close to 
  the unitary line (red line in the 
  Figure~\ref{fig:feiivsoi}).
We run a Kolmogorov-Smirnov (KS) test to 
  verify the similarity of these two populations.
We found a statistical significance of p = 0.74, implying 
  that it is highly likely that both lines belong to the 
  same parent population.
This result can also be observed in 
  Figures~\ref{fig:blrprofiles1} to~\ref{fig:blrprofiles4},
  which show that both lines display similar velocity 
  widths and shapes.
\cite{rod02}, analyzing a smaller sample, found that these 
  two lines had similar profiles, suggesting that they 
  arise from the same parcel of gas.
Our results strongly support these hypothesis using 
  a different and a more sizable sample of 25 AGNs.
   
\begin{figure}
\epsscale{0.85}
\plotone{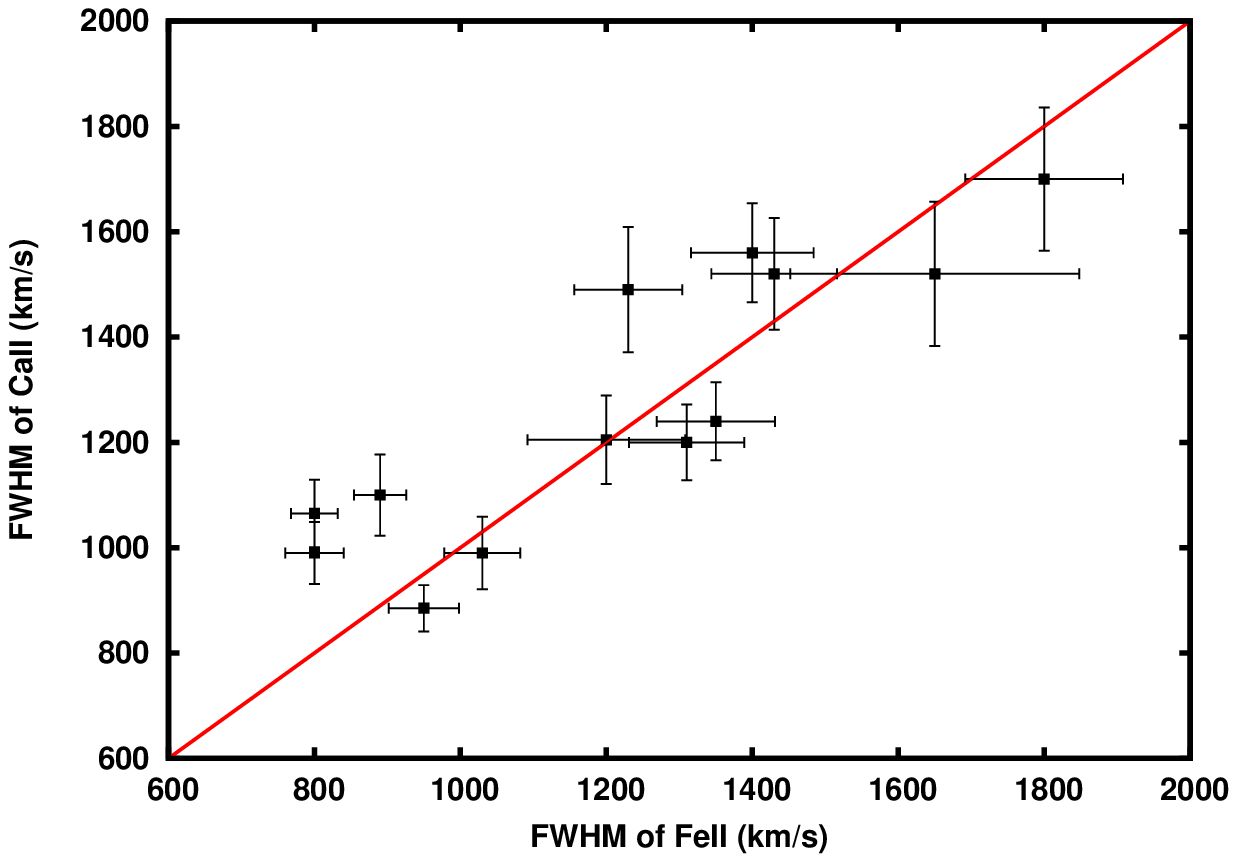}
\caption{Correlation between the FWHM of \feii\ and the FWHM of \caii. The red line shows the unitary relationship.}
\label{fig:feiivscaii}
\end{figure}

A similar behavior is seen in Figure~\ref{fig:feiivscaii}, 
  which shows the FWHM of \caii\ vs \feii. 
The lower number of points is explained by the fact 
  that for a sub-sample of objects it was not possible 
  to obtain a reliable estimate of the FWHM of 
  \caii\ either due to poor S/N or because in some objects 
  \caii\ appears in absorption. 
As with \oi\ and \feii, we found that the width of 
  \caii\ is similar to that of \feii. 
The KS test reveals a statistical significance of p = 0.81.
The combined results of Figures~\ref{fig:feiivsoi} and 
  \ref{fig:feiivscaii} support the physical picture where these 
  three lines are formed in the same region. 
Since the analysis of the \feii\ emission is usually more 
  challenging, the fact of \oi\ and \caii\ are produced 
  co-spatially with iron provides constrains on the use of these 
  ions to study the same physical region of the BLR 
  \citep{mat07,mat08}. 

\begin{figure}
\epsscale{0.85}
\plotone{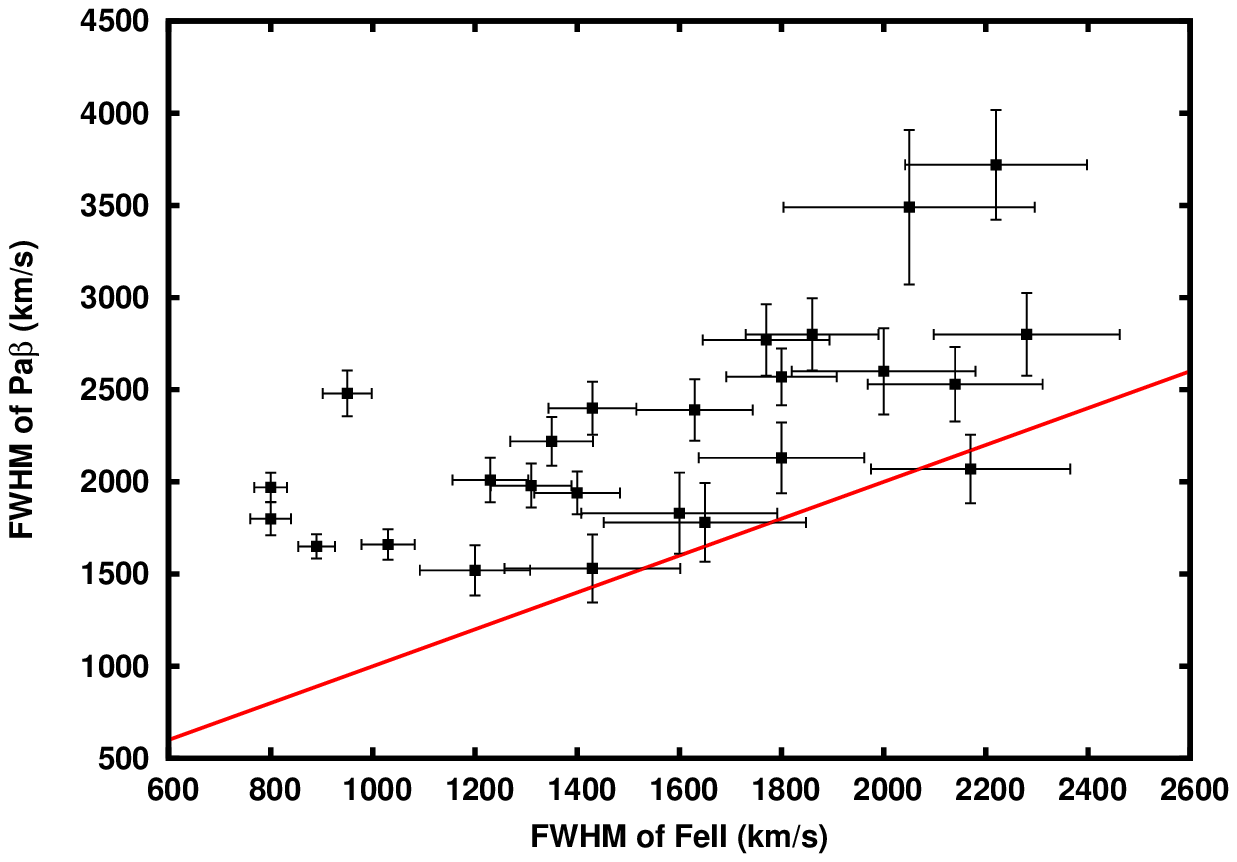}
\caption{Correlation between the FWHM of \feii\ and the FWHM of \pab. The red line shows the unitary relationship.}
\label{fig:feiivspab}
\end{figure}

In contrast to \oi\ and \caii, the Paschen lines display a 
  different behavior. 
Figure~\ref{fig:feiivspab} shows the FWHM of \pab\ 
  vs \feii. 
It is clear that the latter appears systematically 
  narrower than the former, suggesting that the 
  H\,{\sc i} lines are formed in a region closer to the 
  central source than \feii\ and, by extension,
  \oi\ and \caii. 
The KS test for these two populations resulted in an 
  statistical significance p = 0.001.
The average FWHM value for 
  \pab\ is $\sim$30$\%$ larger than that of \feii. 
  
Assuming that the \feii\ emitting clouds are virialized, 
  the distance between the nucleus and the clouds  
  are given by $D\,\propto\,v^{-2}$. 
Using in this 
  equation the average difference in width (or velocity) 
  between \feii\ and H\,{\sc i},
we found that the \feii\ emitting region is twice as far 
  from the nucleus compared to the region where hydrogen emission
  is produced.
  
The stratification of the BLR can also be observed 
  in Figures~\ref{fig:blrprofiles1} to \ref{fig:blrprofiles4}, 
  which compare the line profiles of the above four lines 
  discussed in this Section. 
We add to the different panels, when available, 
  C\,{\sc iv}\,$\lambda$1550, a higher-ionization emission line.
The plots show that \feii, \oi\ and \caii\ 
  have similar FWHM and profile shapes. 
\pab\ has a larger FWHM than \feii, and the 
  C\,{\sc iv} line is usually the broadest of the five lines. 
Moreover, the C\,{\sc iv} line profile is highly asymmetric. 
This result indicates that C\,{\sc iv} is probably emitted in 
  an inner region of the BLR, closer to the central source than 
  \pab\ and is very likely affected by outflows driven 
  by the central source, as well as electron or Rayleigh scattering
  \citep{gas09,laba05}. 
An observational test of this scenario was provided by \cite{nor06}
  who detected P-Cygni profiles in this line in a sample of 
  7 AGNs.
  
The above findings are in good agreement to those reported 
  in the literature for different samples of AGNs and spectral 
  regions.
\cite{hu08} analyzed a sample of more than 4000 
  spectra of quasars from SDSS and verified that the FWHM 
  of the \feii\ lines was, on average, 3/4 that of \hb.
\cite{slu07}, using spectroscopic microlensing studies for 
  the AGN RXS\,J1131-1231, found that \feii\ is 
  emitted most probably in an outer region beyond \hb.
\cite{mat08}, comparing the intensities of 
  \caii/\oi\,$\lambda$8446 and 
  \oi\,$\lambda$11287/\oi\,$\lambda$8446
  with that predicted by theoretical models, found for 11 AGNs 
  that these lines are emitted in the same region of the BLR,
  with common location and gas densities.
\cite{mar15} studied the emission of the 
  \caii\,Triplet + \oi\,8446\,\AA{} in a 
  sample of 14 luminous AGNs with intermediate redshifts
  and found intensities ratios and widths consistent with 
  an outer part of a high density BLR, 
  suggesting these two emission lines could be emitted in 
  regions with similar dynamics.

Recent works based on variability studies indicates that 
  \feii\ and hydrogen are emitted at 
  different spatial locations, with the former being 
  produced farther out than the latter 
  \citep{kue08,kas00,bar13}.
\cite{kue08}, for instance, studied the reverberation 
  behavior of the optical \feii\ lines in Akn\,120. 
They found that the optical \feii\ emission clearly 
  does not originate in the same region as \hb, 
  and that there was evidence of a reverberation 
  response time of 300 days, which implies an origin in 
  a region several times further away from the central 
  source than \hb. 
\cite{bar13} report similar results
  in the Seyfert\,1 galaxies NGC\,4593 and Mrk\,1511 
  and demonstrate that the \feii\ emission in 
  these objects originates in gas located predominantly 
  in the outer portion of the broad-line region 
  (see Table~\ref{tab:distance} for the values). 
\cite{hu15}, however, analysing the reverberation mapping 
  in a sample of 9 AGNS, identified as super-Eddington
  accreting massive black holes (SEAMBH),  
  found no difference between the time lags
  of \feii\ and \hb.

Despite the fact that \feii\ reverberation mapping 
  results are quite rare in the literature, the 
  reverberation of \hb\ is indeed more common 
  \citep{kas00,pet98,pet99}.
\cite{pet98} present a reverberation mapping for 9 AGNs 
  obtained during a 8 year monitoring campaign, where they 
  derived the distance of the \hb\ emitting line region. 
\cite{kas00} collected data from a 7.5 year monitoring 
  campaign in order to determine several fundamental 
  properties of AGNs such as BLR size and black hole masses. 
Four of their objects (Mrk\,335, Mrk\,509, NGC\,4051 
  and NGC\,7469) are common to our work, and they 
  found the distance of the \hb\ emitting 
  region using reverberation mapping.
From the \cite{hu15} reverberation mapped AGNs 
  we identified 3 objects common to our sample
  (Mrk\,335, Mrk\,1044 and Mrk\,493).
From their results, and assuming that the \feii\ emitting clouds 
  are virialized, we may estimate the distance of these iron 
  clouds to the central source using the relation between 
  the measured FWHM of \feii\ and \hb\ for these objects.
The values that we find are presented in Table~\ref{tab:distance}. 

Column 2 and 3 of Table~\ref{tab:distance} show the distance of 
  \hb\ and \feii\ \citep[determined by reverberation mapping]{kas00,hu15}, respectively.
For Mrk\,335, we notice discrepant values for the distance of 
  \hb\ in the work of \cite{hu15} (8.7 light-days) and 
  \cite{kas00} (16.8 light-days).
Column 4 of Table~\ref{tab:distance} shows that our estimations to the 
  distance of the \feii\ emitting line region.
Except for Mrk\,493, our estimations are in good 
  agreement with those measered from reverberation mapping, with 
  a ratio between the distances of \feii\ and \hb\ emitting 
  line region $\sim$2, as we predict.
\cite{hu15} point out that distance of the \feii\ emitting 
  line region maybe related with the intensity the \feii\ emission. 
They noted that the time lags of \feii\ are roughly the same as \hb\ 
  in AGNs with \Rfour\,$>$\,1 (including Mrk\,493 as well the other sources 
  in their sample that are not common to ours), usually 
  classified as strong \feii\ emitters, and longer for those with 
  normal/weak \feii\ emission (\Rfour\,$<$\,1), as seen in
  the sample of \cite{bar13} (and most of our sample).
This indicates that the physical properties of strong \feii\ emitters may be 
  different of the normal \feii\ emitters. 
Observations of this kind 
  of objects are needed to confirm this hypothesis.
  
From the results of \cite{kas00}, \cite{bar13} and \cite{hu15}, 
  we can also estimate a mean distance for the \hb\ and 
  \feii\ emitting line region: $\tau$(\hb)\,=\,19.7 
  light-days and $\tau$(\feii)\,=\,40.5 light-days. 
If we include high luminosity quasars (such as Mrk\,509) 
  the average values are significantly higher: 
  $\tau$(\hb)\,=\,80.3 light-days 
  and $\tau$(\feii)\,=\,164.6 light-days.
 
Assuming a Keplerian velocity field where the BLR 
  emitting clouds are gravitationally bound to 
  the central source, the above results suggest that 
  the low-ionization lines (\feii, \caii\ 
  and \oi) are formed in the same outer region 
  of the BLR. 
Moreover the hydrogen lines would be formed in a region 
  closer to the central source. 
This scenario is compatible with the physical conditions 
  needed for the formation of \feii\ and \oi: 
  that is, neutral gas strongly protected from the incident 
  ionizing radiation coming from the central source. 
These conditions can only be found in the outer regions of 
  the central source. 
Our work confirms results obtained in previous works 
  using different methods in different spectral regions 
  \citep{rod02,per88,hu08,bar13,slu07}, but on 
  significantly smaller samples.

\section{Final Remarks}\label{conclusion}
We analyzed for the first time a NIR sample of 25 AGNs 
  in order to verify the suitability of the NIR \feii\ 
  template developed by \cite{gar12} in measuring the \feii\ 
  strength in a broad range of objects. We also studied the 
  excitation mechanisms that lead to this emission and derived 
  the most likely region where it is produced. 
The analysis and results carried out in the previous 
  sections can be summarized as follows:
\begin{itemize}
 \item We identified, for the first time in a sizable
 sample of AGNs, the \lya\ excitation mechanism 
 predicted by \cite{sipr98}. The key feature of this 
 process is the \feii\ bump at 9200\,\AA{}, which is clearly 
 present in all objects of the sample.
 \item We demonstrated the suitability of the NIR \feii\ 
 template developed by \cite{gar12} in reproducing most 
 of the iron features present in the objects of the sample. 
 The template models and subtracts the NIR \feii\ 
 satisfactorily. We found that the relative intensity of 
 the 1$\mu$m lines remains constant from object to object, 
 suggesting a common excitation mechanism (or mechanisms), 
 most likely collisional excitation. Qualitative analysis 
 made with the NIR and UV spectra lead us to conclude that 
 this process contributes to most of the \feii\ production, 
 but \lya\ fluorescence must also contribute to this 
 emission. However, the percentage of the contribution should 
 vary from source to source, producing the small differences 
 found between the predicted and observed $\lambda$9200 
 bump strengths. Despite this, it is still possible 
 to determine the total \feii\ intensity of the bump.
 \item We found that the NIR \feii\ emission and the 
 optical \feii\ emission are strongly correlated. The strong 
 correlation between the indices \Rone, \Rnine\ and \Rfour\ show 
 that \lya\ fluorescence plays an important role in the 
 production of the \feii\ observed in AGNs.
 \item Through the comparison between the number of \feii\ 
 photons in the 9200\,\AA\ bump and that in the 4570\,\AA\ bump, 
 we determine that \lya\ fluorescence should contribute with
 at least $\sim$18$\%$ to all optical \feii\ flux observed 
 in AGNs. This is a lower limit, since UV spectroscopy at a 
 spectral resolution higher than currently available is needed 
 to estimate the total contribution of this process to the 
 observed \feii\ emission. This result is key to the development 
 of more accurate models that seek to better understand the \feii\ 
 spectrum in AGNs.
 \item The comparison of BLR emission line profiles shows that 
 \feii, \oi\ and \caii\ display similar widths for a given object. 
 This result implies that they all are produced in the 
 same physical region of the BLR. In contrast, the \pab\ profiles 
 are systematically broader than those of iron (30\% broader, on 
 average). This indicates that the former are produced in an 
 region closer to the central source than the latter (2$\times$ 
 closer, on average). These results and data from 
 reverberation mapping allowed us to estimate the distance 
 of the \feii\ emitting clouds from the central source 
 for six objects in our sample. The values found range from a few 
 light-days ($\sim$9 in NGC\,4051) to nearly $\sim$200 (in Mrk\,509). 
 Overall, our results agree with those found independently via 
 reverberation mapping, giving additional support to our approach. 
 These results should also guide us to understand why reverberation 
 mapping has had little success in detecting cross-correlated 
 variations between the AGN continuum and the \feii lines.
\end{itemize}

\acknowledgments

We are grateful to U.S. National Science Foundatation (NSF AST-1409207), 
the Canadean and Brazilean funding agencies (NSERC, FAPEMIG and CNPq) by
their support to this paper. The research presented here use of the
NASA/IPAC Extragalactic Database (NED), which is operated by the Jet
Propulsion Laboratory, California Institute of Technology, under contract with
the National Aeronautics and Space Administration.



\begin{deluxetable}{lccccc}
\tabletypesize{\scriptsize}
\tablecaption{Basic information on the IRTF observations.\label{tab:basicdata}}
\tablewidth{0pt}
\tablehead{
\colhead{AGN} & \colhead{Type} & \colhead{z} & \colhead{Date} & \colhead{Exp. Time(s)} & \colhead{E(B-V)$_G$}
}
\startdata
Mrk\,335	&NLS1	&0.02578 &2000 Oct. 21 	&2400 		&0.030\\
I\,Zw\,1	&NLS1	&0.06114 &2003 Oct. 23	&2400	 	&0.057 \\
Ton\,S180	&NLS1  &0.06198 &2000 Oct. 11	&2400		&0.013 \\
Mrk\,1044	&NLS1  &0.01645 &2000 Oct. 11	&1800		&0.031 \\
Mrk\,1239	&NLS1  &0.01927 &2002 Apr. 21	&1920		&0.065 \\
		&	&	 &2002 Apr. 23	&1920		&	\\
Mrk\,734	&S1	&0.05020 &2002 Apr. 23	&2400		&0.029 \\
PG\,1126-041	&QSO    &0.06000 &2002 Apr. 23	&1920		&0.055 \\
		&	&	 &2002 Apr. 24	&2160		&	\\
H\,1143-182  	&S1    &0.03330 &2002 Apr. 21	&1920		&0.039 \\
NGC\,4051	&NLS1  &0.00234 &2002 Apr. 20	&1560		&0.013 \\
Mrk\,766	&NLS1  &0.01330 &2002 Apr. 21	&1680		&0.020 \\
		&	&	 &2002 Apr. 25	&1080		&	\\
NGC\,4748	&NLS1  &0.01417 &2002 Apr. 21	&1680		&0.052 \\
		&	&	 &2002 Apr. 25	&1440		&	\\
Ton\,156	&QSO    &0.54900 &2002 Apr. 25	&3600		&0.015 \\
PG\,1415+451	&QSO    &0.11400 &2002 Apr. 24	&3960		&0.009 \\
		&	&	 &2002 Apr. 25	&1440		&	\\
Mrk\,684 	&S1    &0.04607 &2002 Apr. 21	&1440		&0.021 \\
Mrk\,478	&NLS1  &0.07760 &2002 Apr. 20	&3240		&0.014 \\
PG\,1448+273	&QSO    &0.06522 &2002 Apr. 24	&2160		&0.029 \\
PG\,1519+226  	&QSO    &0.13700 &2002 Apr. 25	&4000		&0.043 \\
Mrk\,493	&NLS1  &0.03183 &2002 Apr. 20	&1800		&0.025 \\
		&	&	 &2002 Apr. 25	&900		&	\\
PG\,1612+262	&QSO    &0.13096 &2002 Apr. 23	&2520		&0.054 \\
Mrk\,504 	&NLS1  &0.03629 &2002 Apr. 21	&2100		&0.050 \\
1H\,1934-063  	&NLS1  &0.01059 &2004 Jun. 02	&2160		&0.293 \\
Mrk\,509	&S1    &0.34397 &2003 Oct. 23	&1440		&0.057 \\
		&	&	 &2004 Jun. 01	&2160		&	\\
1H\,2107-097  	&S1    &0.02652 &2003 Oct. 23	&1680		&0.233 \\
Ark\,564	&NLS1  &0.02468 &2002 Oct. 10	&1500		&0.060 \\
		&	&	 &2003 Jun. 23	&2160		&	\\
NGC\,7469 	&S1    &0.01632 &2003 Oct. 23	&1920		&0.069 
\enddata
\end{deluxetable}
\begin{deluxetable}{lcc}
\tabletypesize{\scriptsize}
\tablecaption{Optical and UV data obtained from the literature.\label{tab:otherdata}}
\tablewidth{0pt}
\tablehead{
\colhead{AGN} & \colhead{Optical data} & \colhead{UV data}
}
\startdata
Mrk\,335 	& Casleo 	& - 		\\
I\,Zw\,1 	& Casleo 	& HST FOS 	\\
Ton\,S180 	& Casleo 	& HST STIS	\\
Mrk\,1044 	& Casleo 	& HST COS 	\\
Mrk\,1239 	& Casleo 	& 		\\
Mrk\,734 	& KPNO 		& - 		\\
H\,1143-182 	& Casleo 	& - 		\\
NGC\,4748	& Casleo 	& - 		\\
Ton\,156 	& SDSS 		& - 		\\
PG\,1415+451 	& SDSS 		& HST FOS 	\\
Mrk\,478	& KPNO 		& HST FOS	\\
PG\,1448+273	& SDSS 		& - 		\\
PG\,1519+226	& SDSS 		& - 		\\
Mrk\,493 	& SDSS 		& HST FOS 	\\
PG\,1612+262 	& SDSS 		& HST FOS 	\\
1H\,1934-063	& Casleo 	& - 		\\
Mrk\,509	& - 		& HST COS 	\\
1H\,2107-097	& Casleo 	& -		\\
NGC\,7469 	& Casleo 	& -	 	 
\enddata
\end{deluxetable}
\begin{deluxetable}{lcccccccccccc}
\tabletypesize{\scriptsize}
\tablecaption{Measurements of the most relevant BLR features used in this work.  \label{tab:linefit}}
\tablewidth{0pt}
\tablecolumns{19}
\tablehead{
\colhead{} & \multicolumn{2}{c}{\feii\,$\lambda$10502} & 
\multicolumn{2}{c}{\oi\,$\lambda$11297} &  
\multicolumn{2}{c}{\caii\,$\lambda$8663} &
\multicolumn{2}{c}{Pa\,{$\beta$} $\lambda$10502} & \\
\colhead{Object} 
& \colhead{Flux} & \colhead{FWHM} 
& \colhead{Flux} & \colhead{FWHM} 
& \colhead{Flux} & \colhead{FWHM} 
& \colhead{Flux} & \colhead{FWHM}
}
\startdata
Mrk\,335	&	14.4	$\pm$	0.9	&	1230	$\pm$	74	&	26.0	$\pm$	2.3	&	1140	$\pm$	103	&	14.7	$\pm$	1.2	&	1490	$\pm$	119	&	87.1	$\pm$	5.2	&	2010	$\pm$	121	\\
I\,Zw\,1	&	39.8	$\pm$	1.6	&	890	$\pm$	36	&	29.9	$\pm$	2.1	&	820	$\pm$	57	&	28.6	$\pm$	2.0	&	1100	$\pm$	77	&	86.7	$\pm$	3.5$^a$	&	1650	$\pm$	66$^a$	\\
Ton\,S180	&	2.8	$\pm$	0.1	&	1030	$\pm$	52	&	5.5	$\pm$	0.4	&	930	$\pm$	74	&	5.2	$\pm$	0.4	&	990	$\pm$	69	&	24.0	$\pm$	1.2$^a$	&	1660	$\pm$	83$^a$	\\
Mrk\,1044	&	9.7	$\pm$	0.6	&	1480	$\pm$	79	&	11.7	$\pm$	0.7	&	1010	$\pm$	61	&	7.2	$\pm$	0.4	&	1200	$\pm$	72	&	24.2	$\pm$	1.4	&	1800	$\pm$	119	\\
Mrk\,1239	&	29.1	$\pm$	1.7	&	1350	$\pm$	81	&	50.1	$\pm$	4.0	&	1220	$\pm$	98	&	16,2	$\pm$	1.0	&	1240	$\pm$	74	&	135.5	$\pm$	8.1	&	2220	$\pm$	133	\\
Mrk\,734	&	17.8	$\pm$	2.1	&	1600	$\pm$	192	&	17.3	$\pm$	0.9	&	1670	$\pm$	84	&	-		-	&	-		-	&	71.9	$\pm$	8.6$^a$	&	1830	$\pm$	220$^a$	\\
PG\,,1126-041	&	25.0	$\pm$	2.3	&	2000	$\pm$	180	&	39.4	$\pm$	2.4	&	1940	$\pm$	116	&	-		-	&	-		-	&	128.6	$\pm$	11.6$^a$&	2600	$\pm$	234$^a$	\\
H\,1143-182	&	18.5	$\pm$	1.7	&	2170	$\pm$	195	&	34.5	$\pm$	2.1	&	1720	$\pm$	103	&	-		-	&	-		-	&	151.2	$\pm$	13.6	&	2070	$\pm$	186	\\
NGC\,4051	&	20.0	$\pm$	2.4	&	1430	$\pm$	172	&	51.6	$\pm$	2.6	&	1035	$\pm$	52	&	-		-	&	-		-	&	65.1	$\pm$	7.8	&	1530	$\pm$	184	\\
Mrk\,766	&	21.2	$\pm$	2.5	&	1650	$\pm$	198	&	50.9	$\pm$	2.0	&	1380	$\pm$	55	&	17.9	$\pm$	1.6	&	1520	$\pm$	137	&	115.4	$\pm$	13.9	&	1780	$\pm$	214	\\
NGC\,4748	&	21.2	$\pm$	1.9	&	1800	$\pm$	162	&	24.7	$\pm$	2.0	&	1650	$\pm$	132	&	-		-	&	-		-	&	62.9	$\pm$	5.7	&	2130	$\pm$	192	\\
Ton\,156	&	47.8	$\pm$	5.7	&	2050	$\pm$	246	&	38.1	$\pm$	2.3	&	2070	$\pm$	124	&	-		-	&	-		-	&	144.6	$\pm$	17.4	&	3490	$\pm$	419	\\
PG\,1415+451	&	4.4	$\pm$	0.4	&	2140	$\pm$	171	&	5.3	$\pm$	0.3	&	1780	$\pm$	107	&	-		-	&	-		-	&	17.2	$\pm$	1.4$^a$	&	2530	$\pm$	202$^a$	\\
Mrk\,684	&	43.5	$\pm$	2.6	&	1430	$\pm$	86	&	53.0	$\pm$	2.7	&	1560	$\pm$	78	&	64.2	$\pm$	4.5	&	1520	$\pm$	106	&	107.7	$\pm$	6.5$^a$	&	2400	$\pm$	144$^a$	\\
Mrk\,478	&	50.7	$\pm$	3.0	&	1400	$\pm$	84	&	48.1	$\pm$	3.4	&	1300	$\pm$	91	&	40.3	$\pm$	2.4	&	1560	$\pm$	94	&	122.6	$\pm$	7.4$^a$	&	1940	$\pm$	116$^a$	\\
PG\,1448+273	&	10.9	$\pm$	0.5	&	950	$\pm$	48	&	32.3	$\pm$	1.3	&	880	$\pm$	35	&	12.4	$\pm$	0.6	&	885	$\pm$	44	&	66.9	$\pm$	3.3$^a$	&	2480	$\pm$	124$^a$	\\
PG\,1519+226	&	4.4	$\pm$	0.4	&	2280	$\pm$	182	&	7.9	$\pm$	0.9	&	1890	$\pm$	227	&	-		-	&	-		-	&	20.1	$\pm$	1.6	&	2800	$\pm$	224	\\
Mrk\,493	&	16.2	$\pm$	0.6	&	800	$\pm$	32	&	26.1	$\pm$	1.0	&	770	$\pm$	31	&	29.4	$\pm$	1.8	&	1065	$\pm$	64	&	40.0	$\pm$	1.6$^a$	&	1970	$\pm$	79$^a$	\\
PG\,1612+262	&	3.7	$\pm$	0.3	&	1770	$\pm$	124	&	10.9	$\pm$	1.2	&	2310	$\pm$	254	&	-		-	&	-		-	&	63.8	$\pm$	4.5$^a$	&	2770	$\pm$	194$^a$	\\
Mrk\,504	&	6.0	$\pm$	0.4	&	1630	$\pm$	114	&	3.9	$\pm$	0.3	&	1620	$\pm$	130	&	-		-	&	-		-	&	20.5	$\pm$	1.4$^a$	&	2390	$\pm$	167$^a$	\\
1H\,1934-063	&	16.7	$\pm$	1.5	&	1200	$\pm$	108	&	35.7	$\pm$	2.9	&	1000	$\pm$	80	&	28.6	$\pm$	2.0	&	1205	$\pm$	84	&	62.8	$\pm$	5.7	&	1520	$\pm$	137	\\
Mrk\,509	&	287.1	$\pm$	12.6	&	2220	$\pm$	178	&	640.2	$\pm$	16.8	&	2390	$\pm$	239	&	-		-	&	-		-	&	2474.0	$\pm$	250.4$^a$&	3720	$\pm$	298$^a$	\\
1H,\,2107-097	&	29.9	$\pm$	1.8	&	1800	$\pm$	108	&	20.7	$\pm$	1.7	&	1720	$\pm$	138	&	10.8	$\pm$	0.9	&	1700	$\pm$	136	&	116.1	$\pm$	7.0	&	2570	$\pm$	154	\\
Ark\,564	&	17.9	$\pm$	0.9	&	800	$\pm$	40	&	27.8	$\pm$	1.4	&	820	$\pm$	41	&	28.7	$\pm$	1.7	&	990	$\pm$	59	&	56.9	$\pm$	2.8	&	1800	$\pm$	90	\\
NGC\,7469	&	24.3	$\pm$	1.7	&	1860	$\pm$	130	&	48.1	$\pm$	2.9	&	1830	$\pm$	110	&	-		-	&	-		-	&	174.9	$\pm$	12.2	&	2800	$\pm$	196	\\
\enddata
\tablecomments{FWHM in km\,s$^{-1}$. Flux in units of 10$^{-15}$\,erg\,s$^{-1}$\,cm$^{-2}$}. 
\tablenotetext{a}{For these objects, the measurements correspond to \paa, because \pab\ was not available due to the redshift of the source.}
\end{deluxetable}
\begin{deluxetable}{lccccc}
\tabletypesize{\scriptsize}
\tablecaption{Values of the parameters used to convolve the NIR \feii\ template.\label{tab:temppar}}
\tablewidth{0pt}
\tablehead{
\colhead{} & \colhead{} & \colhead{}  & \colhead{} & \colhead{\textit{rms}$^3$} &  \colhead{\textit{rms}$^3$} \\
\colhead{AGN} & \colhead{Flux$^1$} & \colhead{FWHM$^2$} & \colhead{Function} & \colhead{after subtraction}  & \colhead{around 1\,$\mu$ lines}
}
\startdata
Mrk\,335	&	14.1	$\pm$	0.9	&	1220	$\pm$	74	&	Gaussian	&	1.60	&	1.42	\\
I\,Zw\,1	&	38.8	$\pm$	1.6	&	870	$\pm$	36	&	Lorentzian	&	2.60	&	2.40	\\
Ton\,S180	&	2.4	$\pm$	0.1	&	1020	$\pm$	52	&	Gaussian	&	0.29	&	0.35	\\
Mrk\,1044	&	9.0	$\pm$	0.6	&	1330	$\pm$	79	&	Gaussian	&	0.55	&	0.65	\\
Mrk\,1239	&	29.9	$\pm$	1.7	&	1360	$\pm$	81	&	Gaussian	&	2.70	&	2.50	\\
Mrk\,734	&	16.8	$\pm$	2.1	&	1620	$\pm$	192	&	Gaussian	&	1.06	&	1.16	\\
PG\,,1126-041	&	23.9	$\pm$	2.3	&	2040	$\pm$	180	&	Gaussian	&	1.87	&	1.83	\\
H\,1143-182	&	17.9	$\pm$	1.7	&	2150	$\pm$	195	&	Gaussian	&	3.01	&	2.92	\\
NGC\,4051	&	20.8	$\pm$	2.4	&	1420	$\pm$	172	&	Gaussian	&	0.90	&	0.70	\\
Mrk\,766	&	20.8	$\pm$	2.5	&	1620	$\pm$	198	&	Gaussian	&	2.20	&	2.10	\\
NGC\,4748	&	20.2	$\pm$	1.9	&	1780	$\pm$	162	&	Gaussian	&	1.20	&	1.26	\\
Ton\,156	&	46.8	$\pm$	5.7	&	2030	$\pm$	246	&	Gaussian	&	2.20	&	2.03	\\
PG\,1415+451	&	4.1	$\pm$	0.4	&	2110	$\pm$	171	&	Gaussian	&	0.48	&	1.60	\\
Mrk\,684	&	43.1	$\pm$	2.6	&	1420	$\pm$	86	&	Gaussian	&	4.19	&	4.35	\\
Mrk\,478	&	50.3	$\pm$	3.0	&	1400	$\pm$	84	&	Gaussian	&	2.30	&	2.35	\\
PG\,1448+273	&	11.3	$\pm$	0.5	&	920	$\pm$	48	&	Gaussian	&	0.68	&	0.74	\\
PG\,1519+226	&	4.6	$\pm$	0.4	&	2230	$\pm$	182	&	Gaussian	&	6.40	&	5.90	\\
Mrk\,493	&	16.5	$\pm$	0.6	&	800	$\pm$	32	&	Lorentzian	&	2.80	&	2.71	\\
PG\,1612+262	&	3.3	$\pm$	0.3	&	1760	$\pm$	124	&	Gaussian	&	0.56	&	0.48	\\
Mrk\,504	&	6.2	$\pm$	0.4	&	1620	$\pm$	114	&	Gaussian	&	0.69	&	0.76	\\
1H\,1934-063	&	16.2	$\pm$	1.5	&	1200	$\pm$	108	&	Gaussian	&	1.70	&	1.65	\\
Mrk\,509	&	290.0	$\pm$	13.4	&	2250	$\pm$	178	&	Gaussian	&	1.80	&	1.70	\\
1H,\,2107-097	&	29.5	$\pm$	1.8	&	1810	$\pm$	108	&	Gaussian	&	0.89	&	0.84	\\
Ark\,564	&	17.1	$\pm$	0.9	&	810	$\pm$	40	&	Lorentzian	&	2.20	&	2.32	\\
NGC\,7469	&	24.9	$\pm$	1.7	&	1840	$\pm$	130	&	Gaussian	&	1.17	&	1.05	\\
\enddata
\tablenotetext{1}{In units of 10$^{-15}$\,erg\,s$^{-1}$\,cm$^{-2}$\,\AA{}$^{-1}$}
\tablenotetext{2}{In units of km\,s$^{-1}$}
\tablenotetext{3}{In units of 10$^{-17}$\,erg\,s$^{-1}$\,cm$^{-2}$}
\end{deluxetable}
\begin{deluxetable}{lcccccc}
\tabletypesize{\scriptsize}
\tablecaption{Fluxes of the 1$\mu$m \feii\ lines measured with the template.\label{tab:1mulines}}
\tablewidth{0pt}
\tablehead{
\colhead{AGN} & \colhead{9998\,\AA{}} & \colhead{10502\,\AA{}} & \colhead{10863\,\AA{}} & \colhead{11127\,\AA{}} & \colhead{\pab} & \colhead{\Rone}
}
\startdata
Mrk\,335	&	14.8	$\pm$	0.6	&	13.9	$\pm$	0.6	&	10.1	$\pm$	0.4	&	6.8	$\pm$	0.3	&	87.1	$\pm$	5.2	&	0.52	$\pm$	0.07	\\
I\,Zw\,1	&	30.4	$\pm$	1.2	&	29.4	$\pm$	1.2	&	21.9	$\pm$	0.9	&	14.2	$\pm$	0.6	&	52.9	$\pm$	2.1$^a$	&	1.81	$\pm$	0.08	\\
Ton\,S180	&	2.5	$\pm$	0.1	&	2.4	$\pm$	0.1	&	1.8	$\pm$	0.1	&	1.2	$\pm$	0.1	&	14.6	$\pm$	0.7$^a$	&	0.54	$\pm$	0.07	\\
Mrk\,1044	&	8.5	$\pm$	0.3	&	8.0	$\pm$	0.3	&	5.8	$\pm$	0.2	&	4.0	$\pm$	0.2	&	24.2	$\pm$	1.4	&	1.08	$\pm$	0.07	\\
Mrk\,1239	&	29.7	$\pm$	1.2	&	28.1	$\pm$	1.1	&	20.2	$\pm$	0.8	&	13.7	$\pm$	0.5	&	135.6	$\pm$	8.1	&	0.68	$\pm$	0.07	\\
Mrk\,734	&	15.1	$\pm$	0.6	&	16.0	$\pm$	0.6	&	10.7	$\pm$	0.4	&	7.2	$\pm$	0.3	&	43.9	$\pm$	5.3$^a$	&	1.12	$\pm$	0.09	\\
PG\,,1126-041	&	14.5	$\pm$	0.6	&	14.0	$\pm$	0.6	&	10.0	$\pm$	0.4	&	7.0	$\pm$	0.3	&	78.4	$\pm$	7.1$^a$	&	0.58	$\pm$	0.06	\\
H\,1143-182	&	12.0	$\pm$	0.5	&	11.7	$\pm$	0.5	&	8.3	$\pm$	0.3	&	5.8	$\pm$	0.2	&	151.2	$\pm$	13.6	&	0.25	$\pm$	0.02	\\
NGC\,4051	&	20.0	$\pm$	0.8	&	18.8	$\pm$	0.8	&	13.6	$\pm$	0.5	&	9.4	$\pm$	0.4	&	65.1	$\pm$	7.8	&	0.95	$\pm$	0.11	\\
Mrk\,766	&	18.3	$\pm$	0.7	&	17.6	$\pm$	0.7	&	13.1	$\pm$	0.5	&	9.0	$\pm$	0.4	&	115.4	$\pm$	13.9	&	0.50	$\pm$	0.06	\\
NGC\,4748	&	18.0	$\pm$	0.7	&	17.6	$\pm$	0.7	&	12.7	$\pm$	0.5	&	8.8	$\pm$	0.4	&	62.9	$\pm$	5.7	&	0.91	$\pm$	0.07	\\
Ton\,156	&	33.9	$\pm$	1.4	&	30.1	$\pm$	1.2	&	25.0	$\pm$	1.0	&	17.3	$\pm$	0.7	&	144.6	$\pm$	17.4	&	0.74	$\pm$	0.06	\\
PG\,1415+451	&	4.5	$\pm$	0.2	&	4.7	$\pm$	0.2	&	3.2	$\pm$	0.1	&	2.2	$\pm$	0.1	&	10.5	$\pm$	0.8$^a$	&	1.39	$\pm$	0.07	\\
Mrk\,684	&	44.6	$\pm$	1.8	&	43.0	$\pm$	1.7	&	30.9	$\pm$	1.2	&	21.5	$\pm$	0.9	&	65.7	$\pm$	3.9$^a$	&	1.13	$\pm$	0.13	\\
Mrk\,478	&	51.7	$\pm$	2.1	&	51.6	$\pm$	2.1	&	38.7	$\pm$	1.5	&	26.6	$\pm$	1.1	&	74.7	$\pm$	4.5$^a$	&	1.26	$\pm$	0.14	\\
PG\,1448+273	&	10.1	$\pm$	0.4	&	9.9	$\pm$	0.4	&	7.2	$\pm$	0.3	&	4.9	$\pm$	0.2	&	40.8	$\pm$	2.0$^a$	&	0.79	$\pm$	0.07	\\
PG\,1519+226	&	3.7	$\pm$	0.1	&	3.5	$\pm$	0.1	&	2.6	$\pm$	0.1	&	1.8	$\pm$	0.1	&	20.1	$\pm$	1.6	&	0.58	$\pm$	0.06	\\
Mrk\,493	&	11.6	$\pm$	0.5	&	11.2	$\pm$	0.4	&	8.2	$\pm$	0.3	&	5.4	$\pm$	0.2	&	24.4	$\pm$	1.0$^a$	&	1.49	$\pm$	0.09	\\
PG\,1612+262	&	2.2	$\pm$	0.1	&	2.1	$\pm$	0.1	&	1.5	$\pm$	0.1	&	1.0	$\pm$	0.1	&	38.9	$\pm$	2.7$^a$	&	0.18	$\pm$	0.04	\\
Mrk\,504	&	5.3	$\pm$	0.2	&	5.4	$\pm$	0.2	&	3.9	$\pm$	0.2	&	2.6	$\pm$	0.1	&	12.5	$\pm$	0.9$^a$	&	0.49	$\pm$	0.35	\\
1H\,1934-063	&	17.6	$\pm$	0.7	&	16.9	$\pm$	0.7	&	12.2	$\pm$	0.5	&	8.4	$\pm$	0.3	&	62.8	$\pm$	5.7	&	0.88	$\pm$	0.06	\\
Mrk\,509	&	306.1	$\pm$	16.7	&	295.8	$\pm$	10.0	&	211.0	$\pm$	11.7	&	141.1	$\pm$	7.5	&	2474.0	$\pm$	250.4$^a$& 	0.26 	$\pm$	0.10	\\
1H,\,2107-097	&	24.0	$\pm$	1.0	&	23.6	$\pm$	0.9	&	16.7	$\pm$	0.7	&	11.7	$\pm$	0.5	&	116.1	$\pm$	7.0	&	0.65	$\pm$	0.08	\\
Ark\,564	&	11.6	$\pm$	0.5	&	11.3	$\pm$	0.5	&	8.3	$\pm$	0.3	&	5.6	$\pm$	0.2	&	56.9	$\pm$	2.8	&	0.65	$\pm$	0.01	\\
NGC\,7469	&	17.1	$\pm$	0.7	&	16.5	$\pm$	0.7	&	11.9	$\pm$	0.5	&	8.3	$\pm$	0.3	&	174.9	$\pm$	12.2	&	0.31	$\pm$	0.04	\\
\enddata
\tablecomments{FWHM in km\,s$^{-1}$. Flux in units of 10$^{-15}$\,ergs\,s$^{-1}$\,cm$^{-2}$ }. 
\tablenotetext{a}{For these objects, the measurements correspond to \paa, because \pab\ was not available due to the redshift of the source.}
\end{deluxetable}
\begin{deluxetable}{lccccc}
\tabletypesize{\scriptsize}
\tablecaption{Fluxes for the \feii+Pa9 Bump at $\lambda$9200.\label{tab:9200fit}}
\tablewidth{0pt}
\tablehead{
\colhead{AGN} & \colhead{\feii\,$\lambda$9200 bump$^1$} & \colhead{Pa9+fit residuals} & \colhead{Expected Pa9$^2$} & \colhead{Total \feii\ at 9200\,\AA{}} & \colhead{\Rnine} 
}
\startdata
Mrk\,335	&	16.8	$\pm$	0.7	&	45.8	$\pm$	2.7	&	18.2	$\pm$	1.1	&	44.4	$\pm$	2.9	&	0.51	$\pm$	0.03	\\
I\,Zw\,1	&	30.5	$\pm$	1.2	&	27.3	$\pm$	2.3	&	11.1	$\pm$	0.4	&	46.7	$\pm$	3.4	&	0.88	$\pm$	0.04	\\
Ton\,S180	&	2.6	$\pm$	0.1	&	6.0	$\pm$	0.3	&	3.1	$\pm$	0.2	&	5.5	$\pm$	0.3	&	0.33	$\pm$	0.05	\\
Mrk\,1044	&	9.4	$\pm$	0.4	&	8.8	$\pm$	0.5	&	5.0	$\pm$	0.3	&	13.1	$\pm$	0.9	&	0.54	$\pm$	0.03	\\
Mrk\,1239	&	30.9	$\pm$	1.2	&	31.7	$\pm$	1.9	&	28.3	$\pm$	1.7	&	34.3	$\pm$	2.3	&	0.25	$\pm$	0.02	\\
Mrk\,734	&	16.0	$\pm$	0.6	&	18.6	$\pm$	2.2	&	9.2	$\pm$	1.1	&	25.4	$\pm$	3.3	&	0.35	$\pm$	0.04	\\
PG\,,1126-041	&	15.5	$\pm$	0.6	&	43.9	$\pm$	3.9	&	16.5	$\pm$	1.5	&	42.8	$\pm$	4.2	&	0.33	$\pm$	0.03	\\
H\,1143-182	&	12.7	$\pm$	0.5	&	51.1	$\pm$	4.6	&	31.5	$\pm$	2.8	&	32.4	$\pm$	3.2	&	0.21	$\pm$	0.02	\\
NGC\,4051	&	22.7	$\pm$	0.9	&	35.3	$\pm$	4.2	&	13.6	$\pm$	1.6	&	44.5	$\pm$	5.9	&	0.68	$\pm$	0.08	\\
Mrk\,766	&	21.3	$\pm$	0.9	&	53.7	$\pm$	6.4	&	24.0	$\pm$	2.9	&	51.0	$\pm$	6.7	&	0.44	$\pm$	0.05	\\
NGC\,4748	&	20.4	$\pm$	0.8	&	30.1	$\pm$	2.7	&	13.1	$\pm$	1.2	&	37.3	$\pm$	3.7	&	0.59	$\pm$	0.05	\\
Ton\,156	&	38.1	$\pm$	1.5	&	33.1	$\pm$	4.0	&	30.1	$\pm$	3.6	&	41.0	$\pm$	5.4	&	0.28	$\pm$	0.03	\\
PG\,1415+451	&	5.7	$\pm$	0.2	&	7.5	$\pm$	0.6	&	2.2	$\pm$	0.2	&	11.1	$\pm$	1.0	&	0.64	$\pm$	0.05	\\
Mrk\,684	&	50.8	$\pm$	2.0	&	44.2	$\pm$	2.6	&	13.8	$\pm$	0.8	&	81.1	$\pm$	5.4	&	0.75	$\pm$	0.05	\\
Mrk\,478	&	50.5	$\pm$	2.0	&	36.4	$\pm$	2.2	&	15.7	$\pm$	0.9	&	71.2	$\pm$	4.7	&	0.58	$\pm$	0.03	\\
PG\,1448+273	&	11.4	$\pm$	0.5	&	20.1	$\pm$	1.0	&	8.6	$\pm$	0.4	&	22.9	$\pm$	1.3	&	0.34	$\pm$	0.02	\\
PG\,1519+226	&	4.2	$\pm$	0.2	&	7.9	$\pm$	0.6	&	4.2	$\pm$	0.3	&	7.8	$\pm$	0.7	&	0.39	$\pm$	0.03	\\
Mrk\,493	&	13.0	$\pm$	0.5	&	29.0	$\pm$	1.2	&	5.1	$\pm$	0.2	&	36.8	$\pm$	1.6	&	0.92	$\pm$	0.04	\\
PG\,1612+262	&	2.5	$\pm$	0.1	&	12.0	$\pm$	0.8	&	8.2	$\pm$	0.6	&	6.3	$\pm$	0.5	&	0.10	$\pm$	0.01	\\
Mrk\,504	&	6.7	$\pm$	0.3	&	0.1	$\pm$	0.0	&	2.6	$\pm$	0.2	&	4.0	$\pm$	0.3	&	0.20	$\pm$	0.01	\\
1H\,1934-063	&	19.4	$\pm$	0.8	&	31.4	$\pm$	2.8	&	13.1	$\pm$	1.2	&	37.7	$\pm$	3.7	&	0.60	$\pm$	0.05	\\
Mrk\.509	&	336.2	$\pm$	14.8	&	501.3	$\pm$	42.2	&	441.7	$\pm$	60.2	&	395.8	$\pm$	51.7	&	0.16	$\pm$	0.01		\\
1H,\,2107-097	&	26.1	$\pm$	1.0	&	30.1	$\pm$	1.8	&	24.2	$\pm$	1.5	&	32.0	$\pm$	2.1	&	0.28	$\pm$	0.02	\\
Ark\,564	&	13.0	$\pm$	0.5	&	14.3	$\pm$	0.7	&	11.8	$\pm$	0.6	&	15.4	$\pm$	0.8	&	0.27	$\pm$	0.01	\\
NGC\,7469	&	18.1	$\pm$	0.7	&	44.7	$\pm$	3.1	&	36.4	$\pm$	2.6	&	26.3	$\pm$	2.0	&	0.15	$\pm$	0.01	
\enddata
\tablecomments{Fluxes in units of 10$^{-15}$\,erg\,s$^{-1}$\,cm$^{-2}$ }. 
\tablenotetext{1}{Measure from the NIR \feii\ template.}
\tablenotetext{2}{Based on the Paschen decrement.}
\end{deluxetable}
\begin{deluxetable}{lccc}
\tabletypesize{\scriptsize}
\tablecaption{Fluxes for the optical \feii\ and \hb.\label{tab:4570fit}}
\tablewidth{0pt}
\tablehead{
\colhead{AGN} & \colhead{\feii\,$\lambda$4570 bump$^1$} & \colhead{Broad H$\beta^2$} & \colhead{\Rfour}
}
\startdata
Mrk\,335	&	87.5	$\pm$	6.6	&	11.8	$\pm$	0.9	&	0.74	$\pm$	0.11	\\
I\,Zw\,1	&	32.0	$\pm$	1.6	&	1.4	$\pm$	0.1	&	2.32	$\pm$	0.11	\\
Ton\,S180	&	19.7	$\pm$	1.2	&	2.0	$\pm$	0.1	&	1.01	$\pm$	0.14	\\
Mrk\,1044	&	30.7	$\pm$	2.3	&	2.6	$\pm$	0.2	&	1.16	$\pm$	0.11	\\
Mrk\,1239	&	33.5	$\pm$	2.5	&	2.5	$\pm$	0.2	&	1.33	$\pm$	0.20	\\
Mrk\,734	&	14.3	$\pm$	2.1	&	1.2	$\pm$	0.2	&	1.19	$\pm$	0.10	\\
H\,1143-182	&	6.3	$\pm$	0.7	&	1.9	$\pm$	0.2	&	0.34	$\pm$	0.06	\\
NGC\,4748	&	15.0	$\pm$	1.7	&	1.7	$\pm$	0.2	&	0.90	$\pm$	0.12	\\
Ton\,156	&	3.2	$\pm$	0.5	&	0.4	$\pm$	0.1	&	0.86	$\pm$	0.18	\\
PG\,1415+451	&	8.2	$\pm$	0.8	&	0.6	$\pm$	0.1	&	1.47	$\pm$	0.15	\\
Mrk\,478	&	16.1	$\pm$	1.2	&	1.3	$\pm$	0.1	&	1.24	$\pm$	0.08	\\
PG\,1448+273	&	13.2	$\pm$	0.8	&	1.1	$\pm$	0.1	&	1.22	$\pm$	0.12	\\
PG\,1519+226	&	8.5	$\pm$	0.9	&	1.1	$\pm$	0.1	&	0.76	$\pm$	0.16	\\
Mrk\,493	&	19.1	$\pm$	1.0	&	1.1	$\pm$	0.1	&	1.85	$\pm$	0.14	\\
PG\,1612+262	&	6.6	$\pm$	0.6	&	1.5	$\pm$	0.1	&	0.43	$\pm$	0.06	\\
1H\,1934-063	&	37.9	$\pm$	4.3	&	2.7	$\pm$	0.3	&	1.38	$\pm$	0.08	\\
1H,\,2107-097	&	15.7	$\pm$	1.2	&	1.5	$\pm$	0.1	&	1.07	$\pm$	0.10	\\
NGC\,7469	&	4.0	$\pm$	0.4	&	0.5	$\pm$	0.0	&	0.74	$\pm$	0.06	
\enddata
\tablenotetext{1}{Fluxes in units of 10$^{-14}$\,erg\,s$^{-1}$\,cm$^{-2}$}
\tablenotetext{2}{Fluxes in units of 10$^{-13}$\,erg\,s$^{-1}$\,cm$^{-2}$}
\end{deluxetable}
\begin{deluxetable}{lccc}
\tabletypesize{\scriptsize}
\tablecaption{Number of photons for the \feii\ emission.\label{tab:photon}}
\tablewidth{0pt}
\tablehead{
\colhead{} & \colhead{\feii\ photons} & \colhead{\feii\ photons} & \colhead{}\\
\colhead{AGN} & \colhead{in $\lambda$4570 bump$^1$} & \colhead{in $\lambda$9200 bump$^2$} & \colhead{N$_{9200}$/N$_{4570}$ $^3$}
}
\startdata
Mrk\,335	&	20	$\pm$	2	&	10	$\pm$	10	&	0.51	\\
I\,Zw\,1	&	74	$\pm$	8	&	36	$\pm$	4	&	0.48	\\
Ton\,S180	&	46	$\pm$	4	&	30	$\pm$	3	&	0.65	\\
Mrk\,1044	&	71	$\pm$	8	&	31	$\pm$	3	&	0.43	\\
Mrk\,1239	&	77	$\pm$	10	&	16	$\pm$	2	&	0.20	\\
Mrk\,734	&	33	$\pm$	4	&	12	$\pm$	1	&	0.35	\\
H\,1143-182	&	14	$\pm$	2	&	2	$\pm$	1	&	0.10	\\
NGC\,4748	&	35	$\pm$	3	&	17	$\pm$	2	&	0.50	\\
Ton\,156	&	7	$\pm$	1	&	2	$\pm$	1	&	0.25	\\
PG\,1415+451	&	19	$\pm$	2	&	5	$\pm$	1	&	0.27	\\
Mrk\,478	&	37	$\pm$	2	&	33	$\pm$	1	&	0.89	\\
PG\,1448+273	&	30	$\pm$	3	&	11	$\pm$	1	&	0.35	\\
PG\,1519+226	&	20	$\pm$	3	&	4	$\pm$	1	&	0.18	\\
Mrk\,493	&	44	$\pm$	5	&	17	$\pm$	2	&	0.38	\\
PG\,1612+262	&	15	$\pm$	2	&	3	$\pm$	1	&	0.19	\\
1H\,1934-063	&	88	$\pm$	11	&	18	$\pm$	2	&	0.20	\\
1H,\,2107-097	&	36	$\pm$	4	&	15	$\pm$	2	&	0.41	\\
NGC\,7469	&	9	$\pm$	1	&	2	$\pm$	1	&	0.13	
\enddata
\tablecomments{in units of 10$^{13}$.}. 
\tablenotetext{1}{Energy for one photon of \feii\,$\lambda$4570 = 4.33x10$^{-28}$\,erg\,s$^{-1}$\,cm$^{-2}$.}
\tablenotetext{2}{Energy for one photon of \feii\,$\lambda$9200 = 2.15x10$^{-28}$\,erg\,s$^{-1}$\,cm$^{-2}$.}
\tablenotetext{3}{Average N = 0.36}
\end{deluxetable}
\begin{deluxetable}{lccc}
\tabletypesize{\scriptsize}
\tablecaption{Distance of the \hb\ and \feii\ emitting line regions.\label{tab:distance}}
\tablewidth{0pt}
\tablehead{
\colhead{} & \colhead{\hb} & \colhead{\feii} &  \colhead{\feii} \\
\colhead{Object} & \colhead{Reverb. map.} & \colhead{Reverb. map.} & \colhead{Our estimations} 
}
\startdata
Mrk\,335$^{(a)}$	&	8.7$^{+1.6}_{-1.9}$	&	26.8$^{+2.9}_{-2.5}$		&	23.2$^{+2.1}_{-2.1}$	\\
Mrk\,1044$^{(a)}$	&	10.5$^{+3.3}_{-2.5}$	&	13.9$^{+3.4}_{-4.7}$		&	15.6$^{+1.7}_{-1.7}$	\\
Mrk\,509$^{(b)}$	&	79.3$^{+6.3}_{-6.3}$	&			-		&	198.2$^{+34.5}_{-34.5}$	\\
NGC\,4051$^{(b)}$	&	6.5$^{+5.1}_{-5.1}$ 	&			-		&	8.5$^{+6.3}_{-6.3}$	\\
NGC\,7469$^{(b)}$	&	4.9$^{+0.8}_{-0.8}$ 	&			-		&	11.2$^{+1.8}_{-1.8}$	\\
Mrk\,493$^{(a)}$	&	11.6$^{+1.2}_{-2.6}$	&	11.9$^{+3.6}_{-6.5}$		&	49.3$^{+5.6}_{-5.6}$	
\enddata
\tablecomments{Distances in light-days.}
\tablenotetext{a}{Reverberation mapping values from \cite{hu15}.}
\tablenotetext{b}{Reverberation mapping values from \cite{kas00}.}
\end{deluxetable}

\begin{figure}
\epsscale{0.85}
\plotone{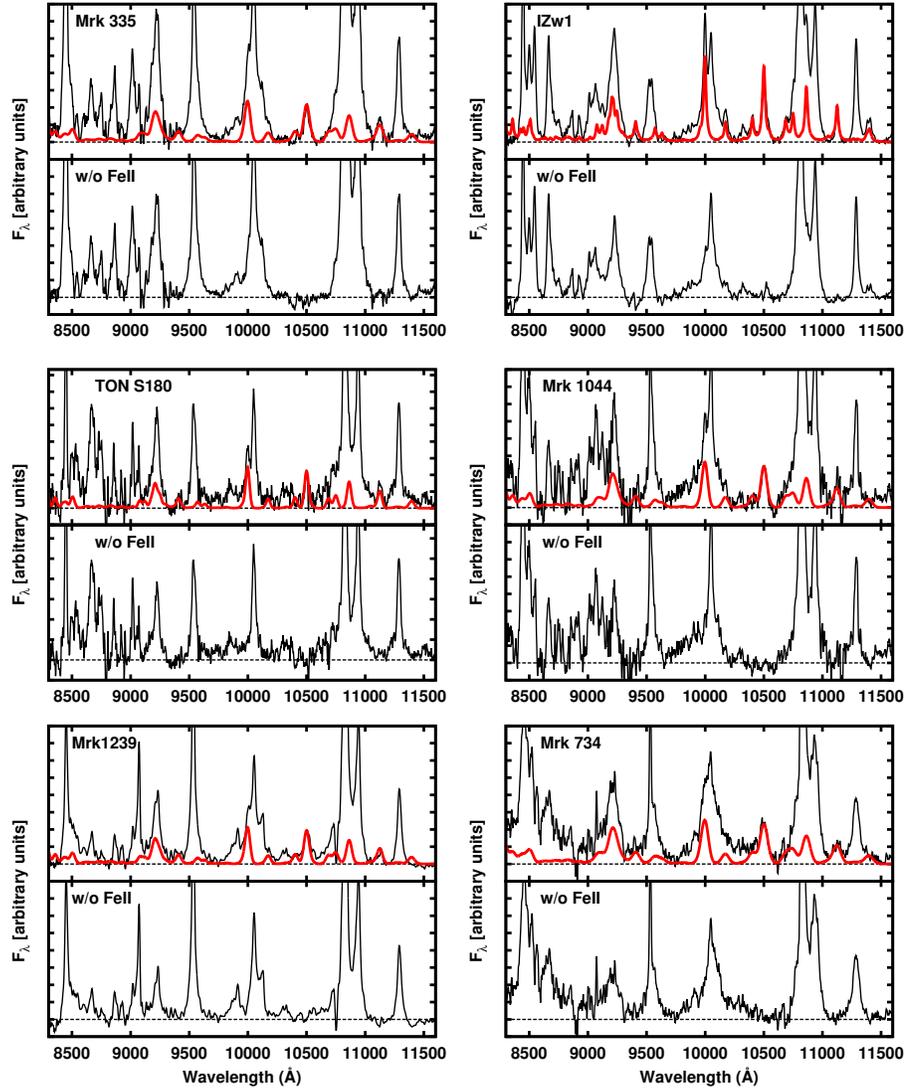}
\caption{Top panels: continuum-subtracted spectrum (in the rest frame) of each object of the 
sample (black line), with the spectrum of \feii\ calculated from the 
semi-empirical template (in bold) superposed. Bottom: spectrum of each object 
of the sample without this contribution.}
\label{fig:nirtemp1}
\end{figure}    
\begin{figure}
\epsscale{0.85}
\plotone{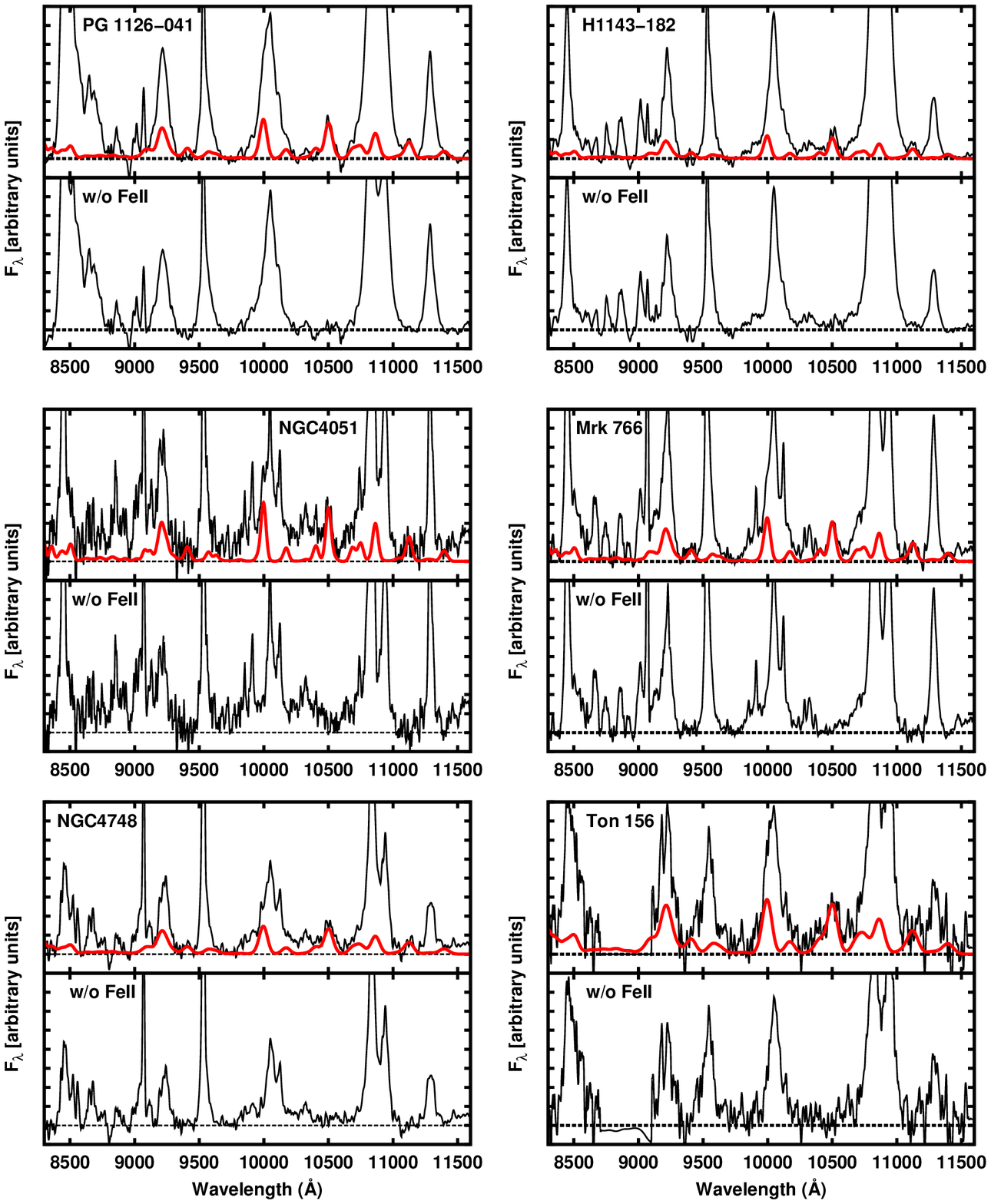}
\caption{Continuation of Figure~\ref{fig:nirtemp1}}
\label{fig:nirtemp2}
\end{figure}    
\begin{figure}
\epsscale{0.85}
\plotone{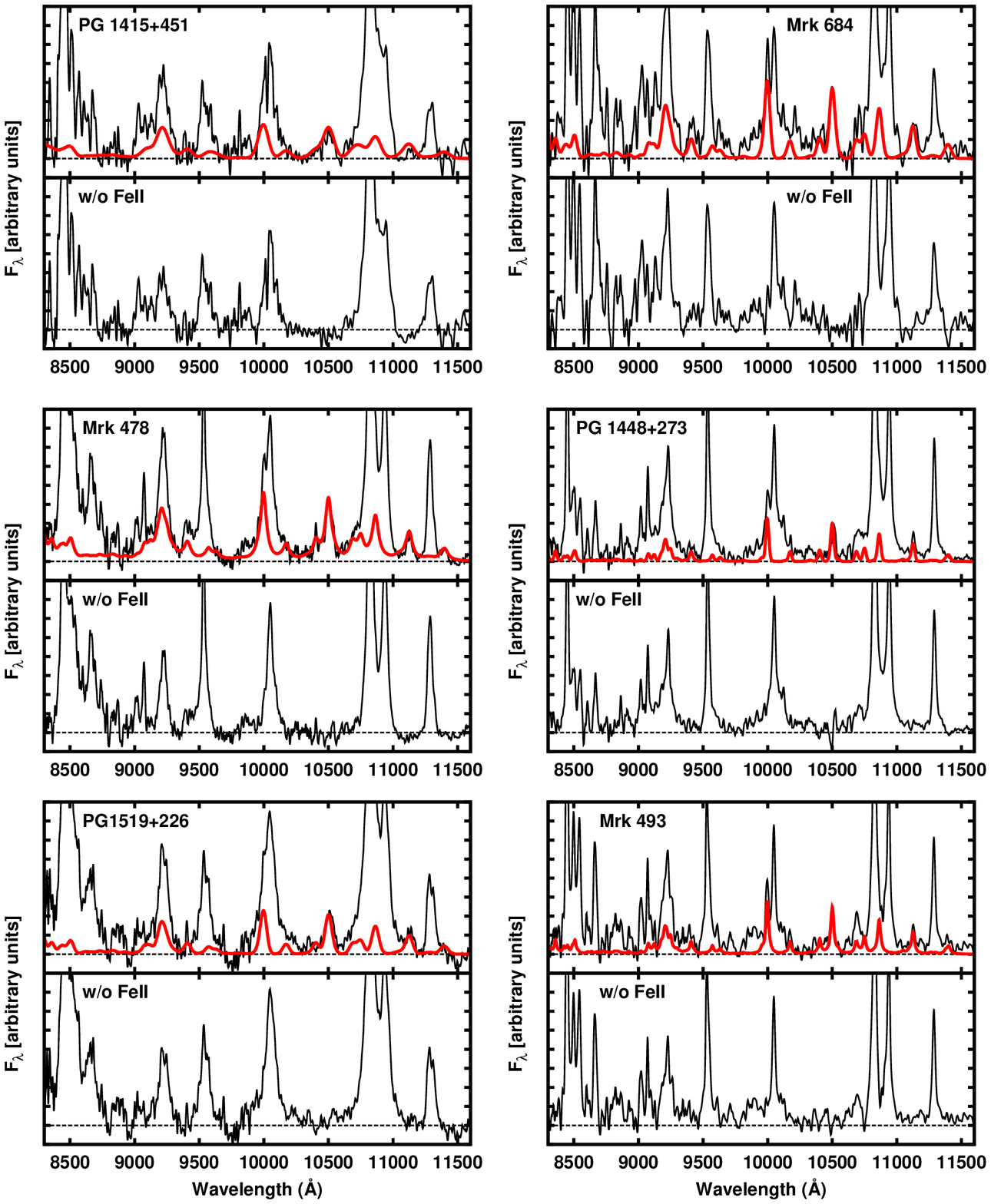}
\caption{Continuation of Figure~\ref{fig:nirtemp1}}
\label{fig:nirtemp3}
\end{figure}    
\begin{figure}
\epsscale{0.85}
\plotone{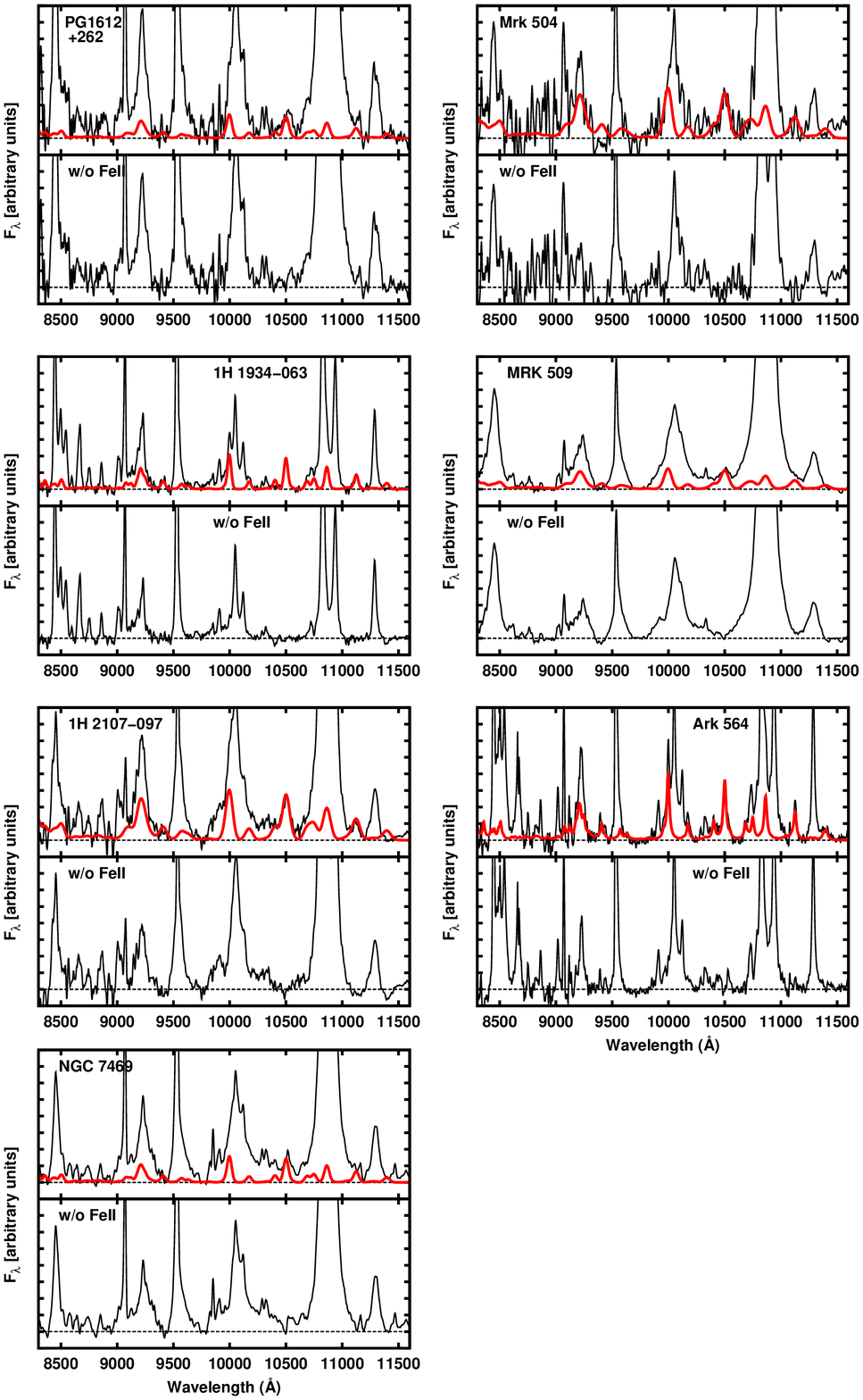}
\caption{Continuation of Figure~\ref{fig:nirtemp1}}
\label{fig:nirtemp4}
\end{figure}    

\clearpage  

\begin{figure}
\epsscale{0.85}
\plotone{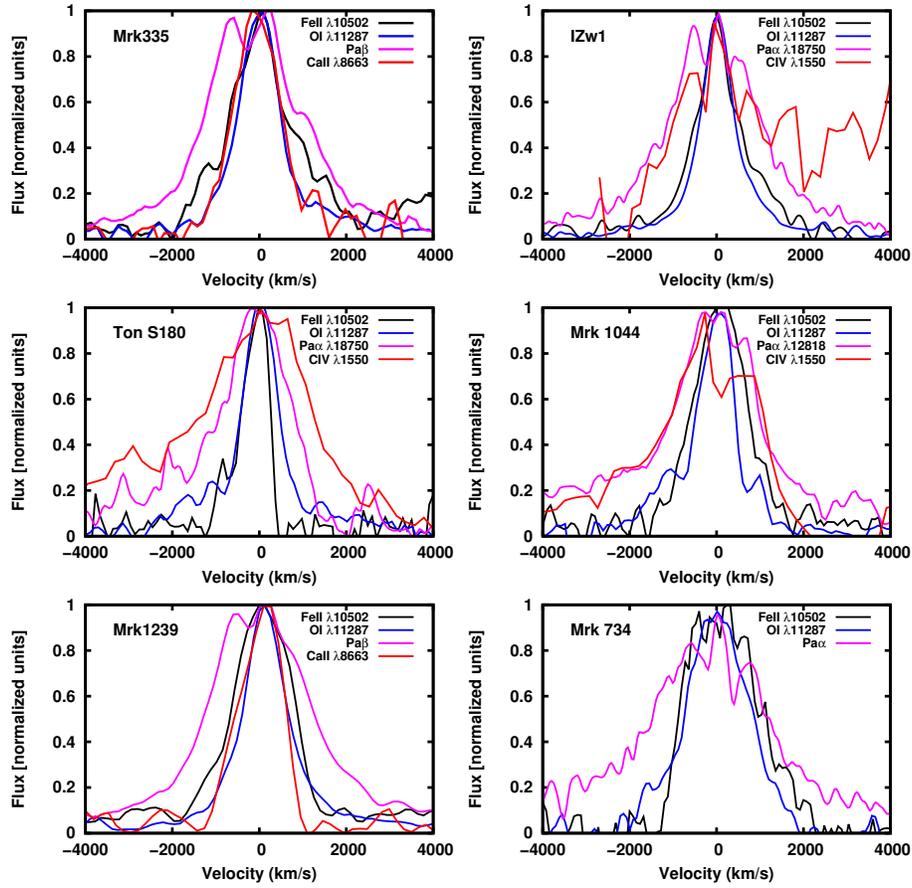}
\caption{Comparison between the broad line profiles for the most important lines in this 
work for each object of the sample: \feii\ (black line), \oi\ (blue line) and \pab\ (magenta line). 
The C\,{\sc iv} (red line) was added for the objects in which it was 
available in order to have a high ionization line to compare.
In all cases, the lines were normalized to their peak intensity and \pab\ had its narrow 
component removed according to the procedure described in Section~\ref{analysis}.}\label{fig:blrprofiles1}
\end{figure}
\begin{figure}
\epsscale{0.85}
\plotone{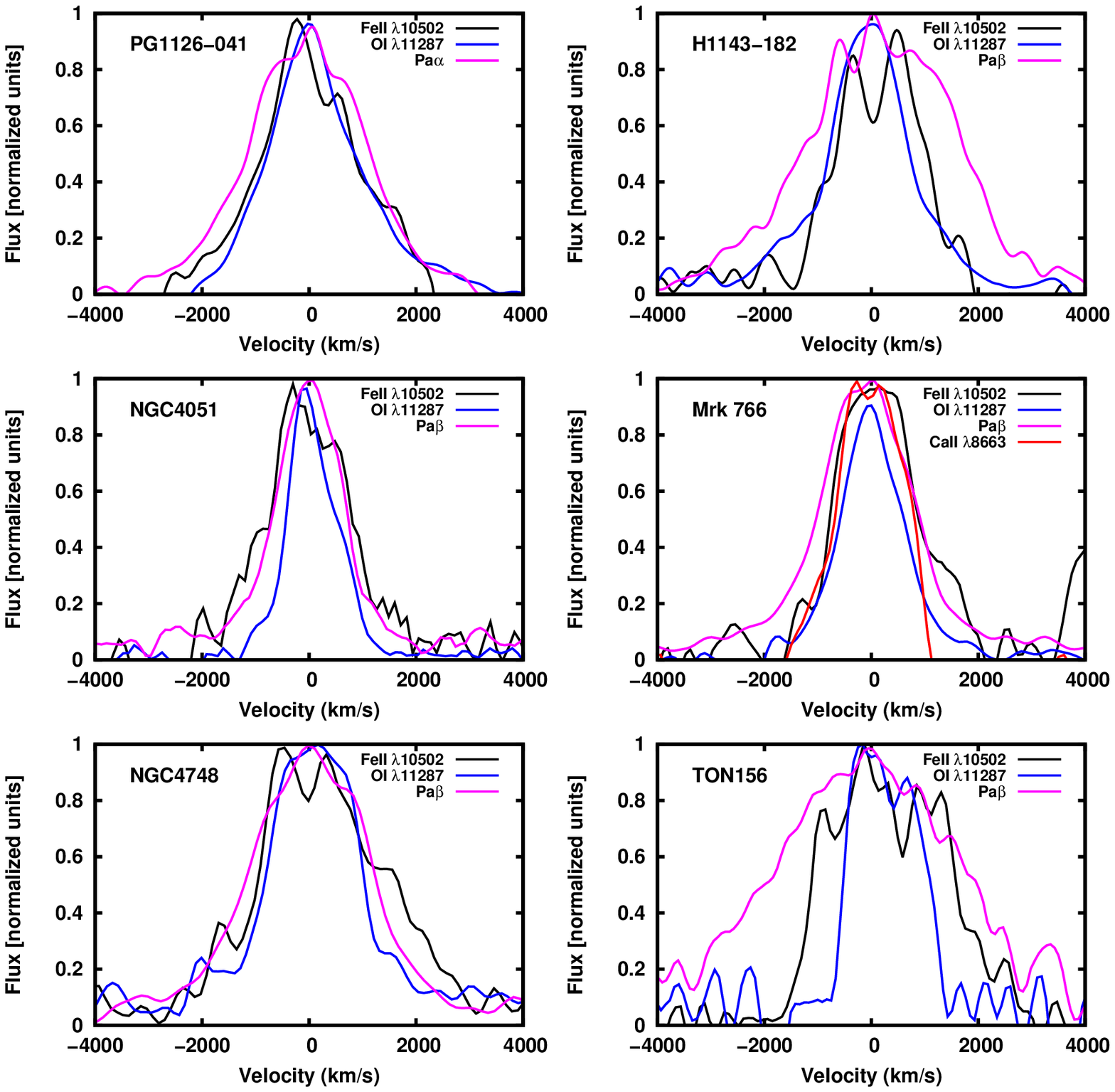}
\caption{Continuation of Figure~\ref{fig:blrprofiles1}}
\label{fig:blrprofiles2}
\end{figure}
\begin{figure}
\epsscale{0.85}
\plotone{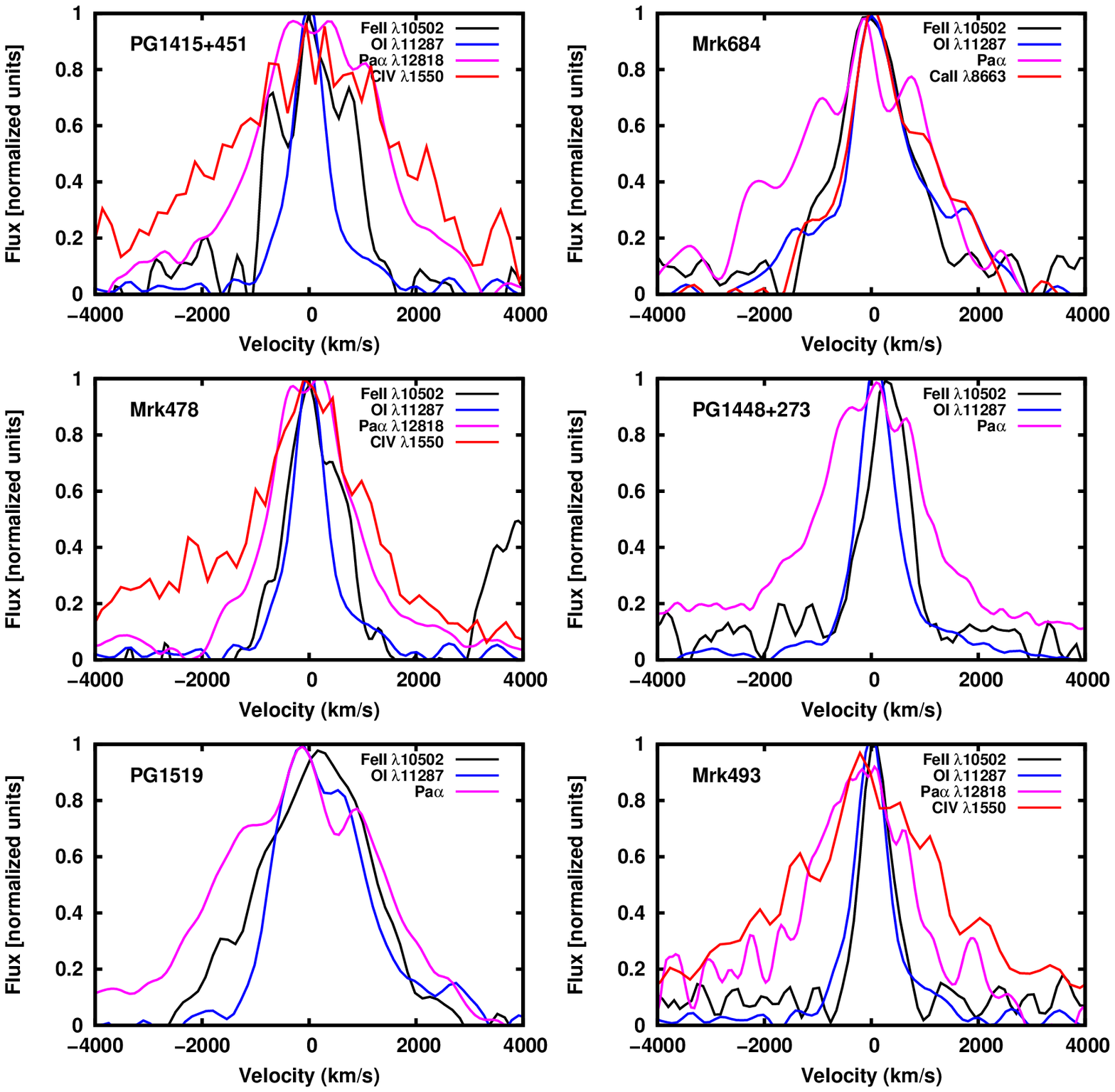}
\caption{Continuation of Figure~\ref{fig:blrprofiles1}}
\label{fig:blrprofiles3}
\end{figure}
\begin{figure}
\epsscale{0.85}
\plotone{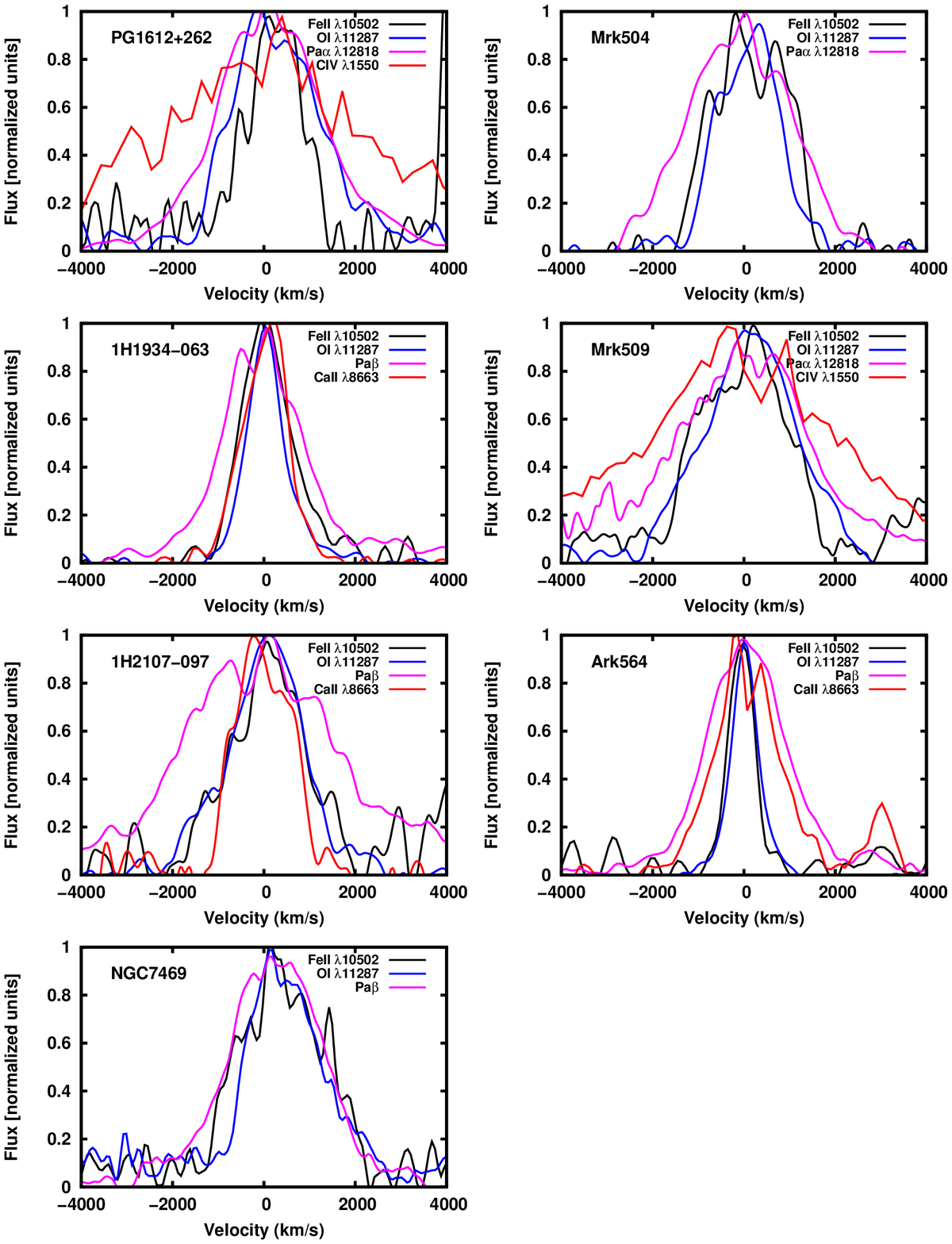}
\caption{Continuation of Figure~\ref{fig:blrprofiles1}}
\label{fig:blrprofiles4}
\end{figure}

\end{document}